\documentclass{aa}  
\usepackage[varg]{txfonts}
\usepackage{graphicx}
\usepackage{txfonts}
\usepackage{float}
\usepackage{lscape}             
\usepackage{placeins}           
                                
\usepackage{hyperref}

\begin{document}

   \title{A negative stellar mass$-$gaseous metallicity gradient relation of dwarf galaxies modulated by stellar feedback}

   \author{Tie Li\inst{1,2}
        \and Hong-Xin Zhang\inst{1,2}\fnmsep\thanks{Corresponding author: hzhang18@ustc.edu.cn}
        \and Wenhe Lyu\inst{1,2}
        \and Yimeng Tang\inst{3}
        \and Yao Yao\inst{1,2}
        \and Enci Wang\inst{1,2}
        \and Yu Rong\inst{1,2}
        \and Guangwen~Chen\inst{1,2,4}
        \and Xu Kong\inst{1,5,6}
        \and Fuyan Bian\inst{7}
        \and Qiusheng Gu\inst{8,9}
        \and Evelyn J. Johnston\inst{10}
        \and Xin Li\inst{8,9}
        \and Shude Mao\inst{11,12}
        \and Yong Shi\inst{8,9}
        \and Junfeng Wang\inst{13}
        \and Xin Wang\inst{14,15,16}
        \and Xiaoling Yu\inst{17}
        \and Zhiyuan Zheng\inst{8,9}
        }

   \institute{Key Laboratory for Research in Galaxies and Cosmology, Department of Astronomy, University of Science and Technology of China, Hefei 230026, China
            \and School of Astronomy and Space Science, University of Science and Technology of China, Hefei 230026, China
            \and Department of Astronomy and Astrophysics, University of California Santa Cruz, 1156 High Street, Santa Cruz, CA 95064, USA
            \and Sub-department of Astrophysics, University of Oxford, Keble Road, Oxford, OX1 3RH, UK
            \and Deep Space Exploration Laboratory/Department of Astronomy, University of Science and Technology of China, Hefei, 230026, People's Republic of China
            \and Frontiers Science Center for Planetary Exploration and Emerging Technologies, University of Science and Technology of China, Hefei, Anhui, 230026, People's Republic of China
            \and European Southern Observatory, Alonso de Cordova 3107, Casilla 19001, Vitacura, Santiago 19, Chile
            \and School of Astronomy and Space Science, Nanjing University, Nanjing 210093, China 
            \and Key Laboratory of Modern Astronomy and Astrophysics (Nanjing University), Ministry of Education, Nanjing 210093, China
            \and Instituto de Estudios Astrof\'isicos, Facultad de Ingenier\'ia y Ciencias, Universidad Diego Portales, Av. Ej\'ercito Libertador 441, Santiago, Chile
            \and Department of Astronomy, Tsinghua University, Beijing, Beijing 100084, China
            \and Department of Astronomy, Westlake University, Hangzhou 310030, Zhejiang Province, China
            \and Department of Astronomy and Institute of Theoretical Physics and Astrophysics, Xiamen University, Xiamen, 361005, China
            \and School of Astronomy and Space Science, University of Chinese Academy of Sciences (UCAS), Beijing 100049, People's Republic of China 
            \and National Astronomical Observatories, Chinese Academy of Sciences, Beijing 100101, People's Republic of China 
            \and Institute for Frontiers in Astronomy and Astrophysics, Beijing Normal University, Beijing 102206, People's Republic of China 
            \and College of Physics and Electronic Engineering, Qujing Normal University, Qujing, Yunnan 655011, China}
   \date{Received November 13, 2024}

  \abstract
   {
   Baryonic cycling is reflected in the spatial distribution of metallicity within galaxies, yet gas-phase metallicity distribution and its connection with other properties of dwarf galaxies are largely unexplored. We present the first systematic study of radial gradients of gas-phase metallicities for a sample of 55 normal nearby star-forming dwarf galaxies (stellar mass $M_{\star}$ ranging from 10$^{7}$ to 10$^{9.5}$ $M_{\odot}$), based on MUSE wide-field spectroscopic observations. We find that metallicity gradient has a significant negative correlation (Spearman's rank correlation coefficient $r$ $\simeq$ $-$0.56) with $M_{\star}$, which is in contrast with the flat or even positive correlation observed for higher-mass galaxies. The negative correlation is accompanied by a stronger central suppression of metallicity compared to the outskirts in lower-mass galaxies. Among the other explored galaxy properties, including baryonic mass, star formation distribution, galaxy environment, regularity of gaseous velocity field and effective yield of metals $\rm y_{eff}$, etc., only the regularity of gaseous velocity field and $\rm y_{eff}$ have residual correlation with metallicity gradient after controlling for $M_{\star}$, in the sense that galaxies with irregular velocity field or lower $\rm y_{eff}$ favor less negative or more positive metallicity gradient. Particularly, a linear combination of logarithmic stellar mass and $\rm y_{eff}$ significantly improves the correlation with metallicity gradients ($r$ $\sim$ $-0.68$) compared to using stellar mass alone. The lack of correlation with environment disproves gas accretion as a relevant factor shaping the metallicity distribution. The correlation with both gaseous velocity field regularity and $\rm y_{eff}$ implies the importance of stellar feedback-driven metal redistribution within the ISM. Our finding suggests that metal mixing and transport process, including but not limited to feedback-driven outflow, are more important than in-situ metal production in shaping the metallicity distribution of dwarf galaxies. 
}
\titlerunning{Metallicity gradients in dwarf galaxies}

\maketitle

\nolinenumbers
\section{Introduction}
Metallicity of the interstellar medium (ISM) serves as a chemical clock of the evolutionary status of its host galaxy. In an idealized closed-box scenario, a galaxy's metallicity is solely determined by the total fraction of gas converted into stars and the stellar yields. But in reality, the metallicity may be subject to modulation by gas accretion from outside or gas outflow driven by feedback processes inside the galaxy \citep[][]{finlator2008,peng2014}. Therefore, ISM metallicity and its spatial distribution can be used to probe galaxy formation history and the accompanied baryon cycle.\par

Studies in the past have found several galaxy scaling relations involving gas metallicity, such as the well-known mass-metallicity (M$-$Z) relation \citep[e.g.,][]{lequeux1979,tremonti2004,yao2022} and the Fundamental Metallicity Relation (FMR) \citep[e.g.,][]{mannucci2010,li2023,bulichi2023}. These studies generally reveal strong positive correlation between metallicity and galaxy mass and negative correlation between metallicity and star formation rate (or gas content) for given galaxy mass. \par

The recent advent and widespread use of Integral Field Unit (IFU) spectroscopy, many large surveys, such as the Calar Alto Legacy Integral Field Area survey (CALIFA, \cite{sanchez2016}), the Mapping Nearby Galaxies at APO (MaNGA, \cite{bundy2015}) and the Sydney-AAO Multi-object Integral-field unit survey (SAMI, \cite{bryant2015}) make it possible to derive relatively accurate 2D distribution of stellar populations, stellar/gaseous kinematics and metallicities for large samples of galaxies. \par

The spatial distributions of gas-phase metallicity in galaxies are usually quantified by the gradient of its radial profiles. Negative gradients of the radial metallicity profiles have been found in most of relatively massive disk galaxies in the local Universe \citep[e.g.,][]{zartsky1994,sanchez2016,pilyugin2014}. This negative gradient is in line with an inside-out formation scenario of the disks \citep[][]{peng2014}, that is, the inner regions of galaxies were first and more efficiently formed by earlier accumulation of gas with low angular momentum (and thus more efficient chemical enrichment), while the gas with higher angular momentum settles in the outer disks at a lower pace and form stars less efficiently. In addition, metal-enriched gas inflow along galaxy disks may also contribute to the negative metallicity gradient \citep[e.g.,][]{wang2022, wang2024}.\par

While studies mentioned above have shown that local spiral galaxies generally have negative gas-phase metallicity gradient, its dependence on various galaxy properties and environment is still not clear. Some recent studies found that gas-phase metallicity gradients are sensitive to galaxy morphology \citep[e.g.,][]{kreckel2019} and environment \citep[e.g.,][]{lara2022}. Actually, a scaling relation between galaxy stellar mass and metallicity gradient is found by some recent work. \cite{belfiore2017} measured the metallicity gradients of a sample of galaxies (log($M_{\star}/M_{\odot}$)>9) from MaNGA survey, and found that galaxies with intermediate stellar masses (log($M_{\star}/M_{\odot}$)$\sim$~10.5) have the steepest metallicity gradients, and the gradients flatten toward both the lower and higher mass end. Similar results are also found in SAMI galaxies \citep[][]{poetrodjojo2021}. Recent simulations \citep[e.g.,][]{hemler2021} and theoretical works \citep[e.g.,][]{sharda2021b} have made significant progress in understanding the physical processes that shape the metallicity gradients. \par

Generally speaking, IFU spectroscopy is superior to traditional long-slit spectroscopy in probing the spatial distribution of metallicity in galaxies. However, the majority of existing large IFU surveys \citep[e.g.,][]{bundy2015,sanchez2016, bryant2015} are biased to relatively massive galaxies. Our current knowledge of metallicity distribution in nearly dwarf galaxies is largely from earlier long-slit spectroscopic observations of selected H II regions. A general conclusion from previous studies is that dwarf galaxies usually lack significant metallicity gradient \citep[e.g.,][]{Pagel1978, Roy1996, Hunter1999,vanZee2006,Lee2007,Croxall2009}. Nevertheless, \cite{pilyugin2015} measured abundance distribution of 14 dwarf irregular galaxies, and found that the metallicity gradient appears to be steeper in galaxies with steeper surface brightness profiles. Due to the faintness and low surface brightness of ordinary star-forming dwarf galaxies, it is observationally expensive to obtain unbiased metallicity distribution for relatively large samples of dwarf galaxies and only several studies attempted to fully explore the mass- and morphology-dependence of metallicity gradients of galaxies by including small sample of dwarf galaxies \citep[e.g.,][]{ho2015, bresolin2019}. More works are needed to extend the relation between the metallicity gradient and other galaxy properties to lower mass end.\par

In this paper, we collect a sample of nearby dwarf galaxies with available MUSE IFU spectroscopic observations, and aim to extend the stellar mass$-$gaseous metallicity gradient relation (MZGR) studies to dwarf galaxy regime, and attempt to explore the physical drivers of the metallicity distribution of dwarf galaxies. This paper is organized as follows. In Sect. \ref{sec:Data reduction}, we describe the sample selection and data reduction. The data analysis is presented in Sect. \ref{sec:Data analysis}. We present our result in Sect. \ref{sec:result} and the discussion in Sect. \ref{sec:discussion}. The summary and conclusion are given in Sect. \ref{sec:summary}. Throughout the paper we assume a $\Lambda$CDM (cold dark matter) cosmology with $H_{0}=70 \mathrm{~km} \mathrm{~s}^{-1} \mathrm{Mpc}^{-1}, \Omega_{\mathrm{m}}=0.3$ and $\Omega_{\Lambda}=0.7$.

\section{Sample and data reduction}
\label{sec:Data reduction}

\subsection{Description of MUSE}
The instrument Multi Unit Spectroscopic Explorer (MUSE; \citep[][]{bacon2010}) is an integral-field spectrograph installed on UT 4 of the Very Large Telescopes (VLT) at the Cerro Paranal Observatory. In its the wide-field mode, MUSE has a field of view (FOV) of $1' \times 1'$ with a spatial sampling of $0.2" \times 0.2"$ (i.e., 0.02 kpc at a distance of 20 Mpc), a spectral sampling of 1.25 \AA{}, and a spectral resolution of 1770 at 465 nm to 3590 at 930 nm. \par

\subsection{Sample selection}
The parent sample of nearby galaxies (< 30 Mpc) with $B$-band absolute magnitude fainter than $-$18.5 mag is retrieved from the extragalactic database Hyperleda\footnote{\url{http://atlas.obs-hp.fr/hyperleda/}}. We then search for wide-field mode MUSE observations of these galaxies in the ESO Science Archive\footnote{\url{http://archive.eso.org/wdb/wdb/eso/muse/form}}. To achieve a reasonable signal-to-noise ratio for emission line detection, we require an on-source exposure time of at least 2000 seconds. To have an adequate spatial coverage for gradient measurement, we further require the MUSE observations cover at least out to one effective radius $R_e$ of a galaxy. It turns out 72 galaxies satisfy the above criteria and their MUSE data are retrieved. As will become clear in Sect. \ref{sec:Data analysis}, 36 of the 72 galaxies are active star-forming galaxies with decent detection of nebular emission lines that can be used to measure the gas-phase metallicity distribution. These 36 galaxies will be used to explore the metallicity gradients in this work.

In addition,  a subset of sources from the Dwarf Galaxy Integral-field Survey (DGIS)\footnote{\url{https://www.dgisteam.com/index.html}} \citep[][]{li2025} is also incorporated into our sample. DGIS aims to acquire observations with spatial resolutions as high as 10 to 100 pc while maintaining reasonably high signal-to-noise ratios with VLT/MUSE and ANU-2.3m/WiFeS. The whole sample is composed of 65 dwarf galaxies with $M_{\star}$ < 10$^{9}$ $M_{\odot}$, selected from the Spitzer Local Volume Legacy Survey. We select the subset of their sample that have MUSE observations (40 sources) and meet our galaxy selection criteria (19 out of 40). This brings the total sample size to 55. We use the datacubes calibrated by DGIS team, but the metallicity gradient and other relevant measurements used in this work are derived by us in a consistent manner with the other galaxies in our sample. The information of these 55 galaxies is given in Table \ref{tab:ana_pro}.

\subsection{MUSE data reduction}
We download the raw science data and calibration files from the archive data center. The data reduction is performed using MUSE pipeline in the EsoReflex environment (\cite{freudling2013}). EsoReflex employs a workflow engine that provides a visual guidance of the data reduction cascade, including the standard processes such as wavelength and flux calibration, sky subtraction, cosmic-ray rejection and combination. Particularly, for sky subtraction, we use dedicated offset sky exposures if available; otherwise we carefully choose sky regions near the edge of the science observations. \par

Before combining the individual exposures calibrated through the pipeline, we perform a visual inspection of spectra extracted from the central regions of individual exposures, in order to find and exclude corrupted observations (due to either poor weather conditions or instrumental failures) from the final combination. 6 of the 72 galaxies are excluded from our sample for this reason, bringing the sample to 66. After combining the valid exposures, we use the Zurich Atmosphere Purge package (ZAP; \cite{soto2016}) to further improve sky subtraction. With the combined spectral cubes in hand, we use the utility ${\rm muse\_cube\_filter}$ in MUSE pipeline to generate broadband images over the SDSS $g,r,i,z$ filters by integrating the data cube in the wavelength direction. These broadband images will be used to improve the flux calibration (see below). \par

\subsection{Refinement of the MUSE flux calibration}
\label{sec:pho}
Flux calibration of the MUSE spectra may be subject to significant uncertainties in a relative and (especially) absolute sense. To remedy this potential problem, we turn to the broadband ($g, r, i, z$) images from the Dark Energy Spectroscopic Instrument (DESI) Legacy Imaging Surveys\footnote{\url{https://www.legacysurvey.org/}}. Because the MUSE field does not always contain isolated bright point sources that are ideal for flux calibration, we decide to perform the calibration by comparing integrated flux over the same central area of our galaxies on the DESI images and the above generated MUSE images. For galaxies with isolated point sources falling in the MUSE field, we also perform the flux calibration with the isolated point sources as a sanity check, and find that the two methods agree with each other within 1 percent. 

We find that the ratios of broadband flux measured from the MUSE and DESI images of our galaxies fall in a narrow range of 0.9--1.1, without significant wavelength dependence. Therefore, we choose to apply the scaling factors derived from the $r$ band calibration of each galaxy to the MUSE data cubes to avoid the influence of residual sky lines and incomplete wavelength coverage of MUSE over the broadband (i.e., $g, z$). Four galaxies (ESO489-G56, NGC2915, UGCA116, UGC3755) in our sample do not have DESI images, so we do not attempt to refine their absolute flux calibration. \par

\section{Data analysis}
\label{sec:Data analysis}
\subsection{Photometric and geometric parameters}
\label{sec:geometrypars}

To prepare for exploring the radial distribution of metallicity and other properties of our sample galaxies, we need to obtain the relevant geometric parameters with broadband images. To this end, we use the elliptical isophote analysis tools in the Python package \textsc{photutils} \citep[][]{photutils2024}. In particular, we determine the galactic center, ellipticity, and major-axis position angle (PA) directly based on isophote fitting to the above generated MUSE $r$-band images, and then use the same ellipse geometry parameters to determine the major-axis effective radius $R_e$ based on the DESI $r$-band images. We note that the MUSE $r$-band images are used to measure all the above-mentioned parameters for the four galaxies without DESI images. One galaxy (NGC1487) is excluded from the sample in this step because it involves a close interaction between three galaxies. The measurements are given in Table \ref{tab:ana_pro}. We have compared our measurements of the geometry parameters with those reported in the HyperLeda database, and confirmed that they are generally consistent with each other within uncertainties (e.g., $\Delta(PA) = -1.7 \pm 34.3$, $\Delta(e) = -0.016 \pm 0.13$).

\subsection{Spectral fitting and emission-line measurement}\label{sec:specfit}
A robust emission-line measurement requires a careful stellar continuum modeling. We follow a workflow similar to the MaNGA Data Analysis pipeline (DAP, \cite{westfall2019}) to analyze our MUSE data cubes. The workflow involves three major parts: a hybrid binning scheme for continuum and emission lines, spectral fitting of the continuum with stellar population models, and emission line measurement on continuum-subtracted spectra.

Before performing the analysis, we mask out the spaxels contaminated by foreground stars or background galaxies, and correct the data cubes for the Galactic extinction by adopting the \cite{schlegel1998} extinction map and the \cite{cardelli1989} extinction law. To perform the continuum modeling, we first use the Voronoi tessellation method (VorBin, \cite{cappellari2003}) to adaptively rebin the spaxels to achieve a minimum continuum S/N of 50 \AA$^{-1}$ (near 5500 \AA{}), and then fit the stacked spectra of each rebinned spaxel with the Penalized Pixel cross-correlation Fitting (pPXF, \cite{cappellari2017}) package. Residual sky lines and nebular emission lines are masked during the continuum fitting, and the fitting is restricted to a wavelength range from 4800 to 7200 \AA{} to avoid contamination from residual sky lines at redder wavelengths and the boundaries of MUSE spectra. We use the E-MILES stellar population models (\cite{vazdekis2016}) and follow the same strategy as \cite{tang2022} to perform pPXF fitting. The reader is referred to \cite{tang2022} for more details. \par

After obtaining the best-fit continuum model for each Voronoi bin, we rescale the continuum model to match that of the observed spectrum of each individual spaxel in the same Voronoi bin, and then subtract the scaled model spectra from individual spaxels. We then estimate S/N and equivalent width (EW) of nebular emission lines, including H$\alpha$, H$\beta$, [O III]$\lambda$5007, [N II]$\lambda$6583, [S II]$\lambda$6716 and [S~II]$\lambda$6731, based on the continuum-subtracted spectra of individual spaxels. 28 galaxies (out of 93) have no emission line detection and are thus excluded from the following analysis. Next, we perform Voronoi binning of the original spaxels to achieve a minimum S/N $\sim$ 10 in the weakest emission lines mentioned above. Spaxels with H$\alpha$ equivalent width < 5 \AA{} may have significant contribution from the so-called Diffuse Ionized Gas (DIG), for which the existing metallicity calibration methods may not be valid (e.g., \cite{Sanders2017}, \cite{zhang2017}). So these low EW(H$\alpha$) spaxels are masked during the Voronoi binning above and excluded from the following analysis.\par

We perform Gaussian profile fitting to each emission line of the Voronoi-binned spaxels to determine the line flux, central velocity and velocity dispersion. To correct for internal dust extinction of the emission lines, we adopt the Balmer decrement method, by assuming an intrinsic H$\alpha$/H$\beta$=2.86 for a Case B recombination with a temperature of 10,000 K and an electron density of 100 $\rm cm^{-3}$ (\cite{hummer1987}). The \cite{cardelli1989} dust-attenuation law is used for the extinction correction. We note that a zero dust extinction is assigned to spaxels with an observed H$\alpha$/H$\beta$<2.86. Lastly, we use the log([OIII]$\lambda$5007/H$\beta$) vs. log([NII]$\lambda$6584/H$\alpha$) Baldwin-Phillips-Terlevich \citep[BPT;][]{baldwin1981}) diagram to exclude spaxels that are inconsistent with being pure star-forming regions, based on the division line of \cite{kauffmann2003} on the BPT diagram. We note that the WHaD method proposed by \cite{Sanchez2024} provides an overall consistent classification of star-forming spaxels with the WH$\alpha$+BPT method adopted in this work, as already pointed out by \cite{Sanchez2024}. After this step, 59 galaxies (out of 65) have at least 10 Voronoi-binned star-forming spaxels. In the end, we re-examined the metallicity distribution in our sample galaxies by visual inspection, and find 4 galaxies (ESO59-01, ESO320-14, ESO321-14, VCC0170) have only metallicity measurements that cover very small radial range or a single star forming region. These 4 galaxies are excluded from the following analysis. The remaining 55 galaxies constitute the final sample to be used in the following analysis. \par

While the main purpose of spectral fitting in this work is to measure nebular emission lines, we also obtain spatial distributions of stellar population properties, such as stellar mass-to-light (M/L) ratio , mass-weighted or light-weighted ages and metallicities, etc. We use the integrated $r$-band M/L from the spectral fitting and the total $r$-band luminosity measured from the DESI images (or the MUSE images for the four galaxies without DESI observations) to estimate the total stellar mass of our galaxies.

\subsection{Star formation rate estimation}
We use the extinction-corrected H$\alpha$ flux to estimate the star formation rate (SFR) for each Voronoi-binned spaxel, assuming a \cite{cappellari2003} IMF. Specifically, we adopt the SFR calibration from \cite{kennicutt1998}: 
\begin{equation}
\label{eq:hasfr}
\textrm{SFR}(M_{\odot}\; \rm yr^{-1})=4.4\times10^{-42}L_{\rm {H\alpha}}(\rm erg\;s^{-1})
\end{equation}
The $\Sigma_{\rm SFR}$ is calculated with correcting the inclination effect. 

\subsection{Measurement of oxygen abundance and radial gradient}
The oxygen abundance (O/H) has been widely used to trace the gas-phase metallicity. A variety of calibrations have been put forward to measure oxygen abundance based on nebular emission lines. Following the recent findings of \cite{easeman2024}, we use the N2S2H$\alpha$ calibration (\cite{dopita2016}) as our default way to estimate the oxygen abundance in this work. This calibration is less dependent on the ionization parameter than most other strong-line calibrations, and given the similar wavelengths of these relevant emission lines, it is insensitive to reddening. Specifically, the calibration is:
\begin{equation}
\label{eq:metal}
	\begin{array}{l}12+\log (\mathrm{O} / \mathrm{H})=8.77+y, \\ y=\log ([\mathrm{N} \text { II}] /[\mathrm{S} \text { II}])+0.264 \times \log ([\mathrm{N} \text { II}] / \mathrm{H} \alpha) \end{array}
\end{equation}

According to \cite{easeman2024}, this calibration has the best overall performance for estimation of oxygen abundance and its gradient, while other strong-line calibrations, such as N2, O3N2 \citep[][]{Marino2013}) and PG16 \citep[][]{Pilyugin2016}, are subject to larger systematic bias. With that being said, we also carried out our analysis based on N2, O3N2 and PG16 methods, and found that, the exact values of abundance and its gradients can vary with calibration methods, but the overall trend discussed throughout this work does not change.\par
To determine the radial abundance gradient, we calculate the deprojected distance $R$ (a.k.a. major-axis distance) of each spaxel as follows
\begin{equation}
\label{eq:deproj}
R=\sqrt{(d\cos{\theta})^2 + (\frac{d\sin{\theta}}{\cos{i}})^2} 
\end{equation}
where $d$ is the projected distance to galactic center directly measured on the images, $\theta$ is the azimuthal angle measured counter-clockwise from the major axis, and $i$ is the disk inclination angle. The inclination angle is derived from the measured minor-to-major axes ratio (b/a) by assuming a constant intrinsic flattening of q = 0.2 \citep[][]{Roychowdhury2013} 
\begin{equation}
\label{eq:inclination}
\cos ^2\mathrm{i}=\frac{(\mathrm{b} / \mathrm{a})^2-\mathrm{q}^2}{1-\mathrm{q}^2}.
\end{equation}

Gradient of oxygen abundance is derived by performing a linear least-squares fitting to 12+log(O/H) distribution as a function of $R$ (i.e., 12+log(O/H) = $\alpha$+$\beta\times R$). Following the common practice of previous studies \citep[e.g.,][]{raj2019}, we normalize the gradients with the $r$-band half-light radius $R_e$. \par

To obtain robust estimation of the gradients and their uncertainties, we randomly re-sample from the valid spaxels in each galaxy with replacement and repeat the linear least-squares fitting for 1000 times. The median and standard deviation of the resultant gradient distribution of each galaxy are taken as the most probable gradient and uncertainty. The same method is adopted to carry out the radial gradient analysis of the logarithmic SFR surface density profiles or specific SFR profiles.

\begin{figure*}
  \centering
  \includegraphics[width=\textwidth]{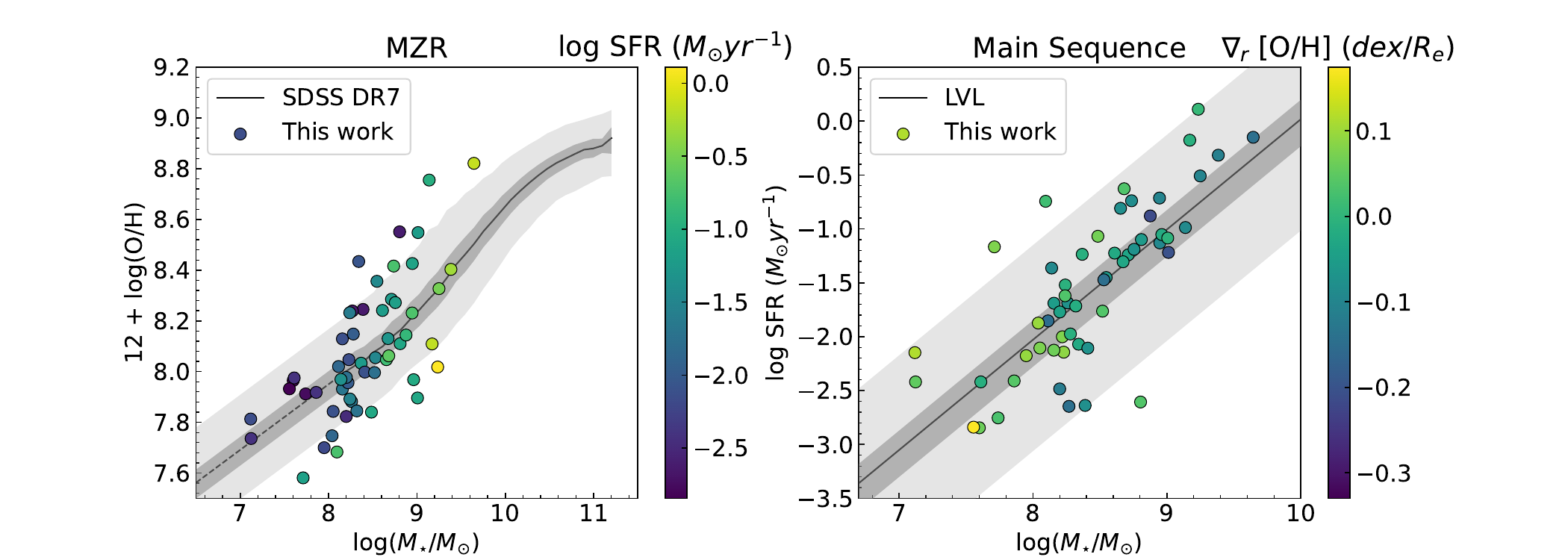}
  \caption{Mass$-$metallicity relation (left panel) and star formation main sequence (right panel) of our galaxy sample (circles color-coded by their SFR in the left panel and by gas-phase metallicity gradient in the right panel, respectively), in comparison with the distribution of SDSS galaxies and LVL galaxies (the solid or dashed lines represent the median trend or linear fit, shaded regions indicates one and two standard deviations from the median trend). The metallicity is derived using N2S2H$\alpha$ method. See Sect. \ref{sec:overview} for details.}
  \label{fig:mzr}
\end{figure*}

\subsection{Estimation of the effective oxygen yield}
\label{sec:eff_measure}
In order to explore the connection between the oxygen abundance gradient and gas inflow / outflow, we make an estimation of the effective oxygen yield (hereafter effective yield) $\rm y_{eff}$ of our galaxies. A galaxy evolving as a closed box obeys a simple analytic relationship between the gas-phase metallicity and the gas mass fraction. As gas is converted into stars, the gas mass fraction decreases and the metallicity of the gas increases according to \cite{searle1972} as:
\begin{equation}
\label{eq:ytrue}
	Z_{\rm gas}= \rm y_{true}{\ln (1 / \mu)}
\end{equation}
where $\rm y_{true}$ is the true nucleosynthetic yield, defined as the ratio of the mass of heavy elements returned to the interstellar medium and the total mass converted to stars but not returned to ISM. If a galaxy evolves as a closed box, the ratio of Eq. \ref{eq:ytrue} should be a constant equal to the nucleosynthetic yield. However, this ratio will be lower if metals have been lost from the system through outflows of gas, or if the gas content has been diluted with fresh infall of metal-poor gas. To quantify the deviation from the closed-box chemical evolution model, the effective yield has been defined as:
\begin{equation}
\label{eq:yeff}
	\rm y_{eff}=\frac{Z_{\rm gas}}{\ln (1 / \mu)}
\end{equation}
where Z$_{\rm gas }$ is the mass fraction of oxygen in gas, and $\mu$ is gas mass fraction with respect to the total of gas and stars. The effective yield would be constant ($\rm y_{true} = \rm y_{eff}$) for a galaxy that has evolved as a closed box. A significant deviation of $\rm y_{eff}$ from the true stellar yield would signify either gas inflow or outflow. Analytical chemical evolution models \citep[e.g.,][]{Kudritzki2015} have clearly shown that galaxies experiencing either outflows/galactic winds or metal-poor gas inflows attain lower metallicities for a given observed gas mass fraction. Such outflow or inflow process is reflected by a lower $\rm y_{eff}$ than the true stellar yield. Therefore, the effective yield serves as a valuable observational metric for diagnosing the processes of gas accretion and removal in galaxies.

The metallicity $Z_{\text {gas }}$ is equal to 12$\times$(O/H), where (O/H) is the number ratio of oxygen and hydrogen atoms. The gas mass includes contribution from both atomic and molecular gas. We collect single-dish HI gas mass $M_{\rm HI}$ measurements from, in order of preference, ALFALFA \citep{haynes2018}, HIPASS \citep{meyer2004}, the All Digital HI Catalog in the Extragalactic Distance Database \citep{Courtois2009} and \cite{loni2021}. Three galaxies (CGCG007-025, NGC1522, PGC132213) in our sample do not have HI observations and one galaxy (FCC119) was not detected in HI emission. NGC4809A and NGC4809B, classified as two galaxies in the early stage of interaction, only have HI measurement for the whole system. For these 6 galaxies we do not calculate their $\rm y_{eff}$. There is no molecular gas observation for most of our galaxies, so we adopt an indirect way to estimate the molecular gas mass by inverting the observationally established molecular gas$-$SFR relation of nearby galaxies. Specifically, we adopt the \cite{kennicutt1998} relation:
\begin{equation}
    {\rm SFR}(M_{\odot}\;yr^{-1}) = 1.4M_{\rm H_2}/10^9
\end{equation}
where SFR is estimated from H$\alpha$, as described above. Taking into account the contribution of Helium and metals, the total gas mass is $\simeq$ 1.35$\times(M_{\rm HI} + M_{\rm H_{2}})$. Estimation of the stellar mass is as described in Sect. \ref{sec:specfit}.

\section{Results}
\label{sec:result}

\subsection{Overview of the galaxy sample}
\label{sec:overview}

Before delving into the metallicity gradient, it is necessary to give an introduction to the global properties of our sample. We start from the well-established galaxy stellar mass$-$gas metallicity relation (MZR) and star formation main sequence (SFMS). In this step, we sum up the relevant emission line fluxes from all valid SF spaxels of each galaxy and estimate the global gas metallicity through Eq. \ref{eq:metal} and SFR through Eq. \ref{eq:hasfr}. The global gas metallicities of the sample are listed in Table \ref{tab:ana_pro}, and the mass$-$metallicity distribution (MZR) is shown in the left panel of Fig. \ref{fig:mzr}, together with the star formation main sequence (SFMS) in the right panel.\par

As a comparison, in the left panel of the figure we also show the MZR of local universe galaxies drawn from the SDSS-DR7 \citep[][]{asplund2009}. Emission-line fluxes of the SDSS galaxies are taken from the OSSY catalog \footnote{\url{https://data.kasi.re.kr/vo/OSSY/index.html}} \citep[][]{oh2011}, and we require a S/N > 3 for all the emission lines used for metallicity estimates, as for our sample. As a result, we are left with about 110,000 galaxies. The metallicities of these SDSS galaxies are estimated with the same method as our sample galaxies. Instead of plotting the individual SDSS galaxies in Fig. \ref{fig:mzr}, we calculate median and standard deviation (with 3-$\sigma$ clipping) for galaxies falling into different logarithmic stellar mass bins from $10^8$ to $10^{11} M_{\odot}$. In the low galaxy stellar mass end where the incompleteness of SDSS sample becomes significant, we perform a linear fitting to the median MZR of $10^{8.2}$ to $10^9 M_{\odot}$ and extrapolate the best-fit relation down to $10^{6.5} M_{\odot}$ in stellar mass. For the main sequence, we compare our measurements with the main sequence trends at $z\sim$ 0 from the local volume legacy (LVL) survey \citep[][]{Dale2023}. We perform a linear fitting to their sample of stellar mass under $10^{10} M_{\odot}$. The comparison shown in Fig. \ref{fig:mzr} suggests that our sample galaxies largely follow the MZR and SFMS relation, with a possible exception for the few most massive galaxies  ($>$ 10$^{9}$ $M_{\odot}$) that have systematically higher metallicities and SFR than the median trends.

\begin{figure*}
  \centering
  \includegraphics[width=\textwidth]{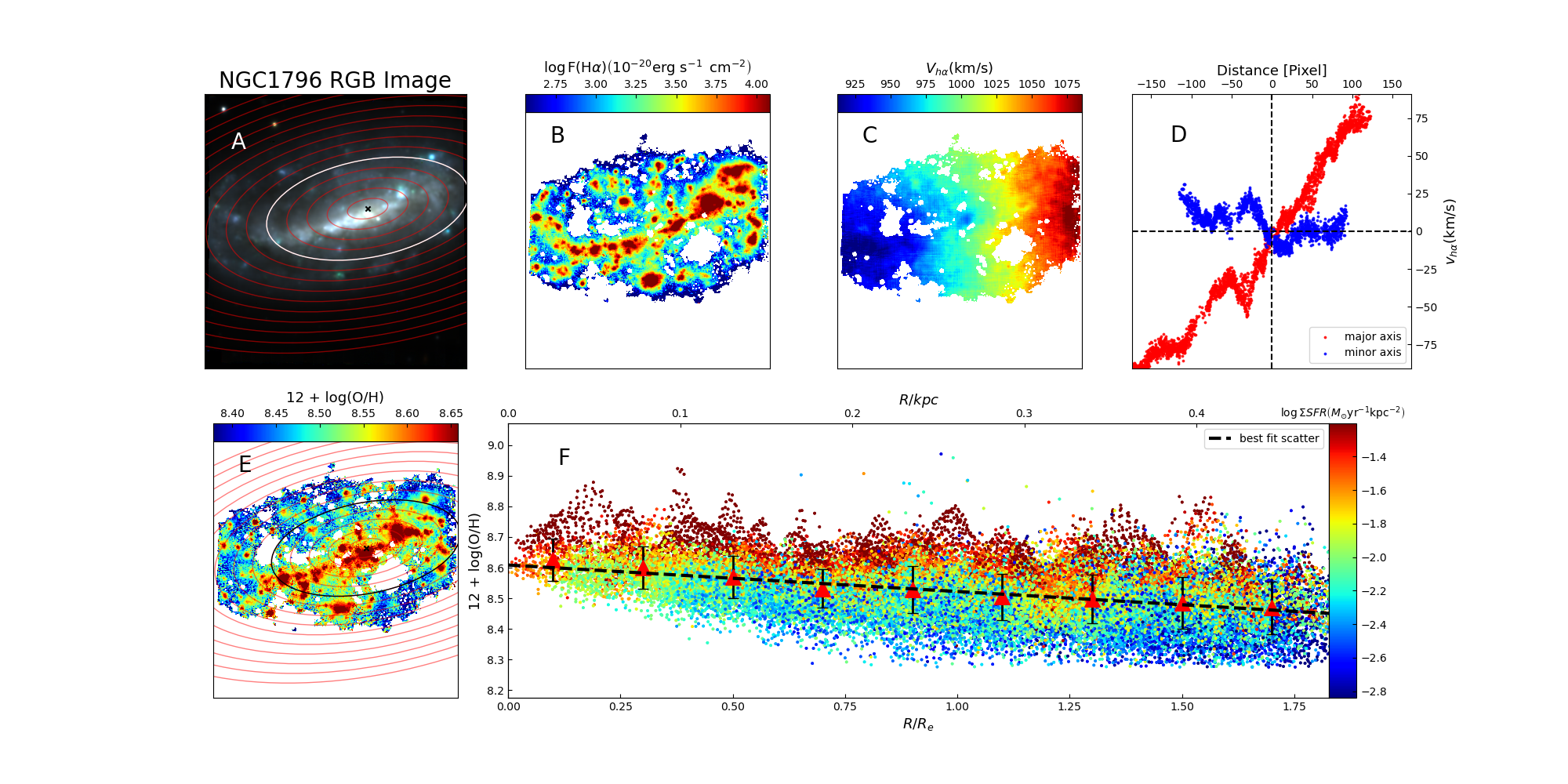}
  \includegraphics[width=\textwidth]{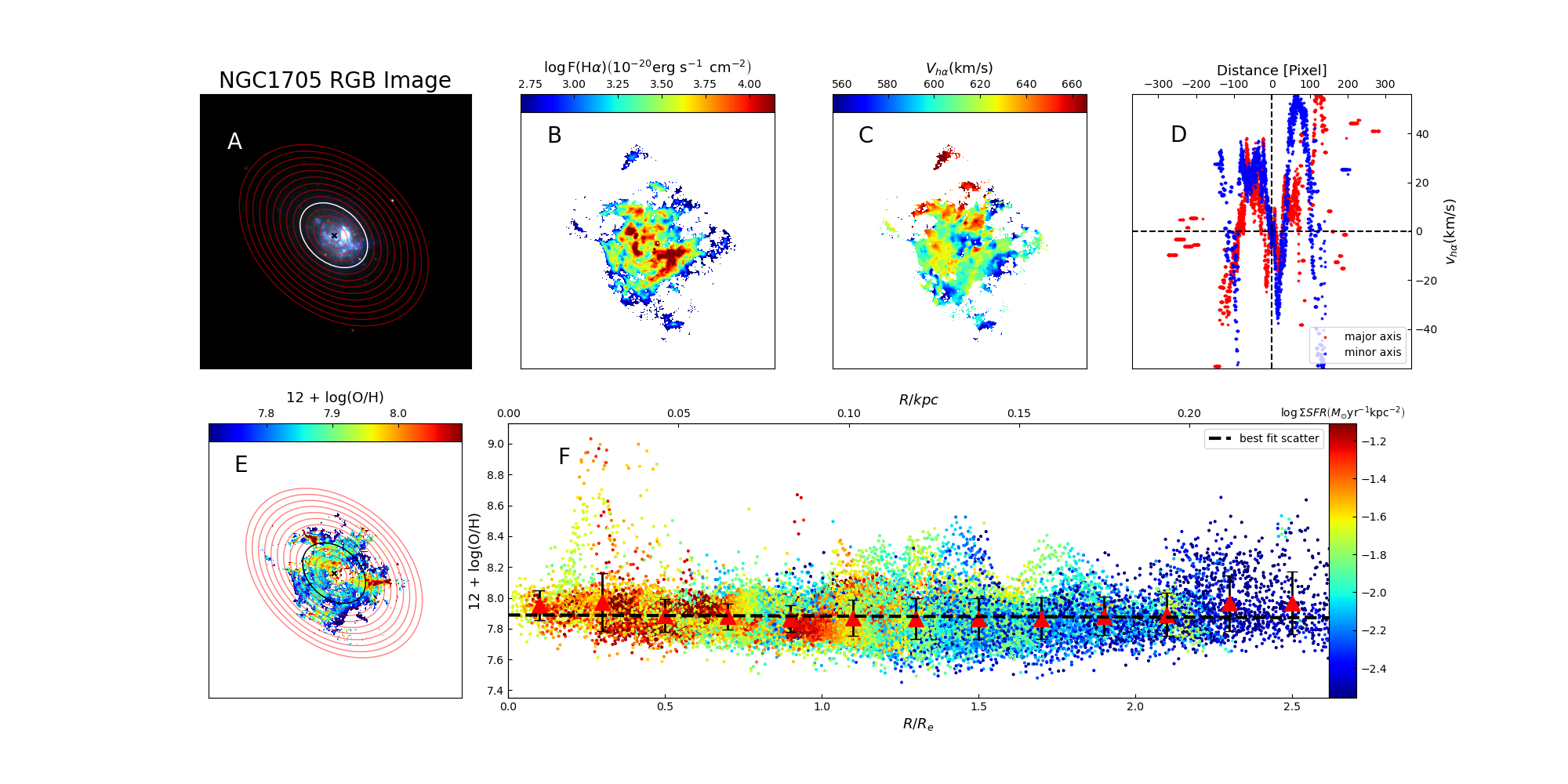}
  \caption{The galaxies NGC1796 and NGC1705 as examples of our working sample. From top left to bottom right for each galaxy: the three-color composite image (A) based on synthetic V, I, and R band images obtained from the MUSE cubes, over-plotted with the isophotal ellipses of galaxy in 0.2 $\rm R_e$ interval (the white ellipse represents 1$\rm R_e$), the H$\alpha$ emission-line map (B) and its velocity field (C). The H$\alpha$ velocity along the major or minor axis (D), the 12+$\log$(O/H) map (E) and its distribution as a function of galactocentric radius (F). In panel F, the data points are color coded by SFR, and the red triangles are the median values for each 0.2 $R_e$ interval, with error bars denoting the standard deviation. The dashed line corresponds to the best-fit linear relation of the data points. Similar figures for the other galaxies in our sample can be found at \protect\url{https://github.com/Count-Lee/metallicity-gradient}.}
  \label{fig:metal_measure}
\end{figure*}

\subsection{Spatial distribution and radial profiles of metallicities}
\label{sec:metal_measure}

In this section we present the resolved maps of gas-phase metallicity, SFR(H$\alpha$) and emission line velocities. The radial gradients of metallicity and SFR are listed in Table \ref{tab:ana_pro}. \par

As examples, the resolved maps and radial profiles of NGC1796 and NGC1705 are shown in Fig. \ref{fig:metal_measure}. NGC1796 is chosen as a typical galaxy in our sample with clear negative metallicity gradient, while NGC1705 serves as an example with no significant metallicity gradient.

NGC1796 (upper panels of Fig. \ref{fig:metal_measure}) has an obvious velocity gradient, suggesting a regular rotating disk. Subplot F of the upper panels of Fig. \ref{fig:metal_measure} suggests that, at given radius, regions with higher SFR tend to also have higher metallicity. This positive correlation between SFR and metallicity at small scale, is consistent with the result of \cite{wang2021}. They proposed that this reflects a changing star formation efficiency rather than a changing gas inflow rate. NGC1705 (lower panels of Fig. \ref{fig:metal_measure}) does not have an obvious H$\alpha$ velocity gradient, suggesting a significant disturbance of the velocity field. We note that an irregular gaseous velocity field does not necessarily mean a lack of disk rotation, as the gas component can be easily disturbed by outflow and inflow activities. Unlike NGC1796, there is no clear positive correlation between local SFR and metallicities in NGC 1705 (Subplot E of the lower panel). This may reflect a stochastic spatial distribution of star formation activities or an efficient metal mixing in the galaxy.

Among our sample, 18 galaxies have a clear positive metallicity gradient, while the rest have either a negative or zero gradient. With a logarithmic median stellar mass of 8.36 (in $M_{\odot}$), the median metallicity gradient of our sample is $-$0.021 $\pm$ 0.84, with a typical median uncertainty of 0.016 in unit of dex kpc$^{-1}$, and $-$0.026 $\pm$ 0.092 with median uncertainty of 0.018 in unit of dex $R_{\rm e}^{-1}$. The median gradient is much shallower than typical high-mass galaxies. For example, a median gradient of $-$0.08 dex kpc$^{-1}$ is found for MaNGA galaxies in the stellar mass range 9.0 < log($M_{\star}/M_{\odot}$) < 11.5 \citep[][]{belfiore2017}. It is our goal in this work to explore the physical drivers of the shallow or inverted metallicity gradients of dwarf galaxies. \par

\subsection{Mass-Metallicity gradient relation}
\label{sec:mzgr}

\begin{figure}
  \centering
  \includegraphics[width=\linewidth]{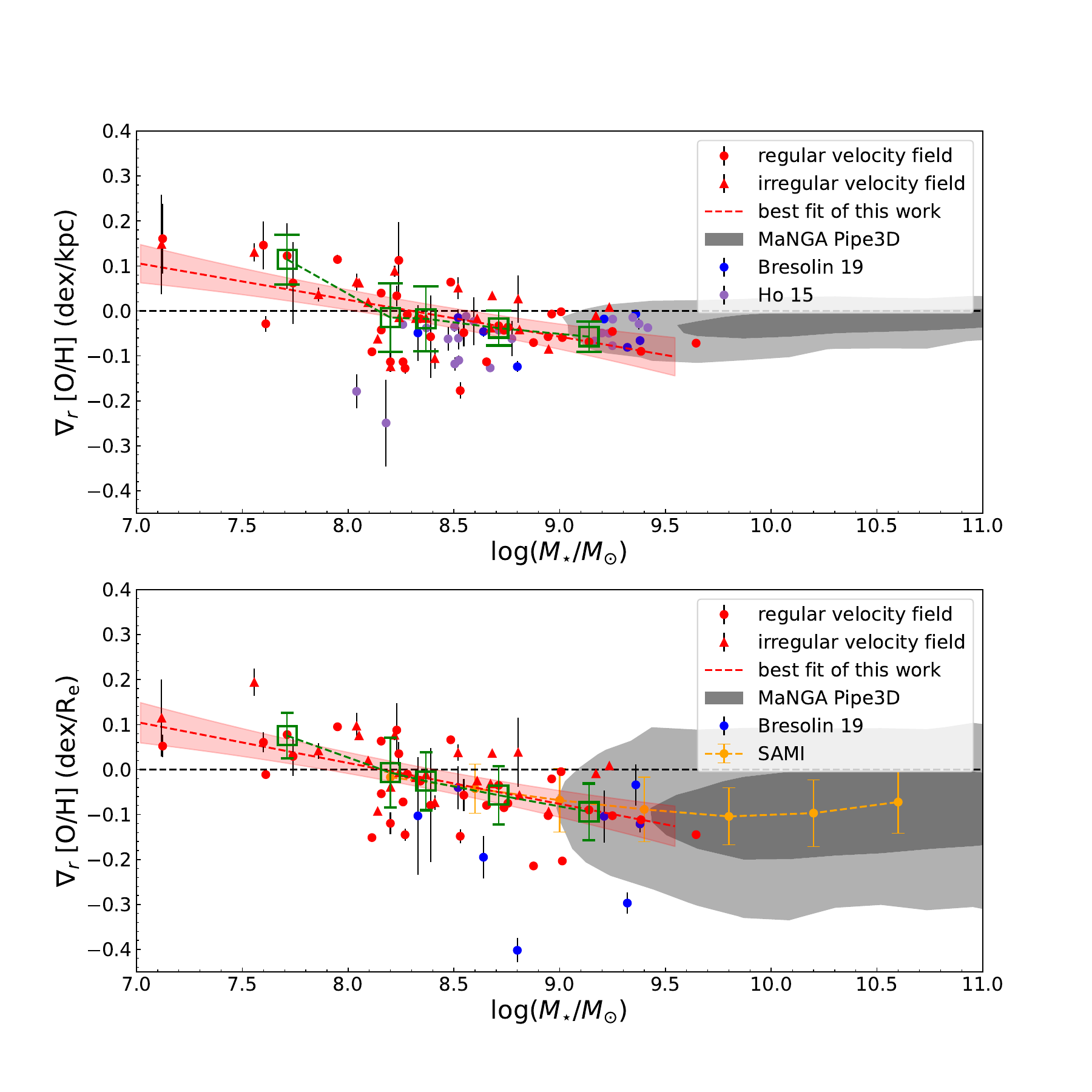}
  \caption{Metallicity gradient versus galaxy stellar mass relation, where the metallicity gradient is expressed in dex kpc$^{-1}$ in the upper panel and in dex $R_{\rm e}^{-1}$ in the lower panel. The red circles and triangles represent the galaxies with or without regular velocity field measured in this work. The red dashed lines and shaded region represent the best fit and scatter of our sample. The green squares show the median values of metallicity gradient and their 1 $\sigma$ dispersion as a function of galaxy stellar mass. Some of the results from previous studies \citep[][]{ho2015,bresolin2019,sanchez2018,poetrodjojo2021} are also shown for comparison. See Sect. \ref{sec:mzgr} for details.}
  \label{fig:mzgr}
\end{figure}

\begin{table*}
\caption{Linear fit and Spearman's rank correlation test for the relationship between the metallicity gradient and other properties.}
\centering
\setlength{\tabcolsep}{5mm}{
\begin{tabular}{lllllr}
\hline
property & unit & slope& intercept & r & p-value \\
\hline

$\log \rm M_{\star}$            &dex kpc$^{-1}$           &$-$0.081$\pm$0.016    &0.679$\pm$0.142       &$-$0.47$\pm$0.089     &0.000249  \\
$\log \rm M_{\star}$            &dex $R_{\rm e}^{-1}$     &$-$0.091$\pm$0.017    &0.743$\pm$0.149       &$-$0.56$\pm$0.081     &0.000093  \\
$\log \rm M_{baryon}$           &dex kpc$^{-1}$           &$-$0.081$\pm$0.018    &0.705$\pm$0.158       &$-$0.49$\pm$0.091     &0.00013   \\
$\log \rm M_{baryon}$           &dex $R_{\rm e}^{-1}$     &$-$0.089$\pm$0.017    &0.855$\pm$0.159       &$-$0.58$\pm$0.073     &0.0000032 \\
$\nabla_{r} \log$ SFR           &dex $R_{\rm e}^{-1}$     &$-$0.016$\pm$0.009    &$-$0.036$\pm$0.016    &$-$0.13$\pm$0.12      &0.36      \\
$\nabla_{r} \log$ sSFR          &dex $R_{\rm e}^{-1}$     &$-$0.011$\pm$0.018    &$-$0.024$\pm$0.011    &$-$0.11$\pm$0.13      &0.43      \\
$\log \rm y_{eff}$              &dex $R_{\rm e}^{-1}$     &$-$0.13$\pm$0.034     &$-$0.36$\pm$0.091     &$-$0.37$\pm$0.10      &0.0061    \\
$\Phi$                          &dex $R_{\rm e}^{-1}$     &$-$0.09$\pm$0.019     &0.8$\pm$0.164         &$-$0.46$\pm$0.095     &0.00038   \\
$\nabla_{r}\log\Sigma_{\star}$  &dex $R_{\rm e}^{-1}$     &0.062$\pm$0.035       &0.026$\pm$0.032       &0.13$\pm$0.14         &0.33      \\
$\eta_k$                        &dex $R_{\rm e}^{-1}$     &0.005$\pm$0.008       &$-$0.03$\pm$0.017     &$-$0.10$\pm$0.14      &0.47      \\

\hline
\end{tabular}}
\tablefoot{Col(1)-(2): Galaxy properties and units. Col(3)-(4): The slope and intercept of linear fit between the metallicity gradient and other properties with errors estimated through bootstrap random sampling. Col(5)-(6): The Spearman's rank correlation coefficient $r$ and p-values. A p-value below 0.05 indicates the correlation coefficient is statistically significant.}
\label{tab:fit}
\end{table*}

The galaxy stellar mass$-$metallicity gradient relation of our dwarf galaxies is shown in Fig. \ref{fig:mzgr}. For comparison purposes, literature samples of low-mass galaxies \citep[][]{ho2015,bresolin2019} and high-mass galaxies from MaNGA Pipe3D Value-added catalog \citep[][]{sanchez2016} are also plotted. The MaNGA galaxies are presented as the gray shaded region in Fig. \ref{fig:mzgr}. The \cite{ho2015} sample includes metallicity gradients of galaxies in units of dex kpc$^{-1}$ and dex $R_{25}^{-1}$, covering a stellar mass range of $10^8M_{\odot}$ to $10^{11}M_{\odot}$, with 21 galaxies below a stellar mass of $10^{9.5}M_{\odot}$ and 14 galaxies below $10^9M_{\odot}$. Their sample galaxies are represented as purple dots. \cite{bresolin2019} collects a sample of small and nearby spiral galaxies, and their metallicity gradients are based on long-slit spectroscopy of H II regions. Here we only plot the 8 galaxies (blue dots) with stellar mass lower than $10^{9.5}M_{\odot}$ in their sample. We also include the metallicity gradients from the SAMI survey \citep{poggianti2017} for comparison. SAMI observed a number of low-mass galaxies and calculated their metallicity gradients in units of dex $R_{\rm e}^{-1}$  \citep[][]{poetrodjojo2021}. Here we only plot the median trend of SAMI results for clarity. 

The upper panel of Fig. \ref{fig:mzgr} shows the metallicity gradients in units of dex kpc$^{-1}$, and the lower panel shows the gradients normalized to $R_{\rm e}^{-1}$. To aid in visualizing the mass-dependent behavior, we group our galaxies into five mass intervals (11 galaxies each) and calculated the median values and standard deviation of their metallicity gradients. The results are shown as green squares in Fig. \ref{fig:mzgr}. The median gradients decrease with increasing stellar mass across the mass range explored here, no matter how the gradients are quantified. To quantify the strength of the correlation, we calculate the Spearman's rank correlation coefficient $r$. The derived $r$ values are in the range of $\sim$ $-0.5$ to $-0.6$ (Table \ref{tab:fit}), with the gradients normalized by $R_{\rm e}$ having a more negative $r$ (stronger negative correlation). This suggests a moderate correlation between the metallicity gradients and stellar mass. The p-values are $\lesssim$ 10$^{-5}$, far below the 0.05 threshold, rejecting the null hypothesis that the correlation arises by chance.

Given the limited sample size, we fit a simple linear relation between metallicity gradients and log($M_{\star}$) for our sample:
\begin{equation}
\label{eq:mzgr}
\nabla_{r}[\mathrm{O} / \mathrm{H}]=\mathrm{\alpha} \log M_{\star}+\mathrm{\beta}
\end{equation}
The best-fit linear relation is shown as red dashed line and the uncertainty is represented by the shaded region in Fig. \ref{fig:mzgr}). Our finding that the correlation becomes stronger when the radius is normalized by $R_{e}$ for dwarf galaxies aligns with previous studies of higher-mass galaxies \citep[e.g.,][]{belfiore2017,sharda2021b}.\par

\begin{figure*}
  \centering
  \includegraphics[width=0.95\textwidth]{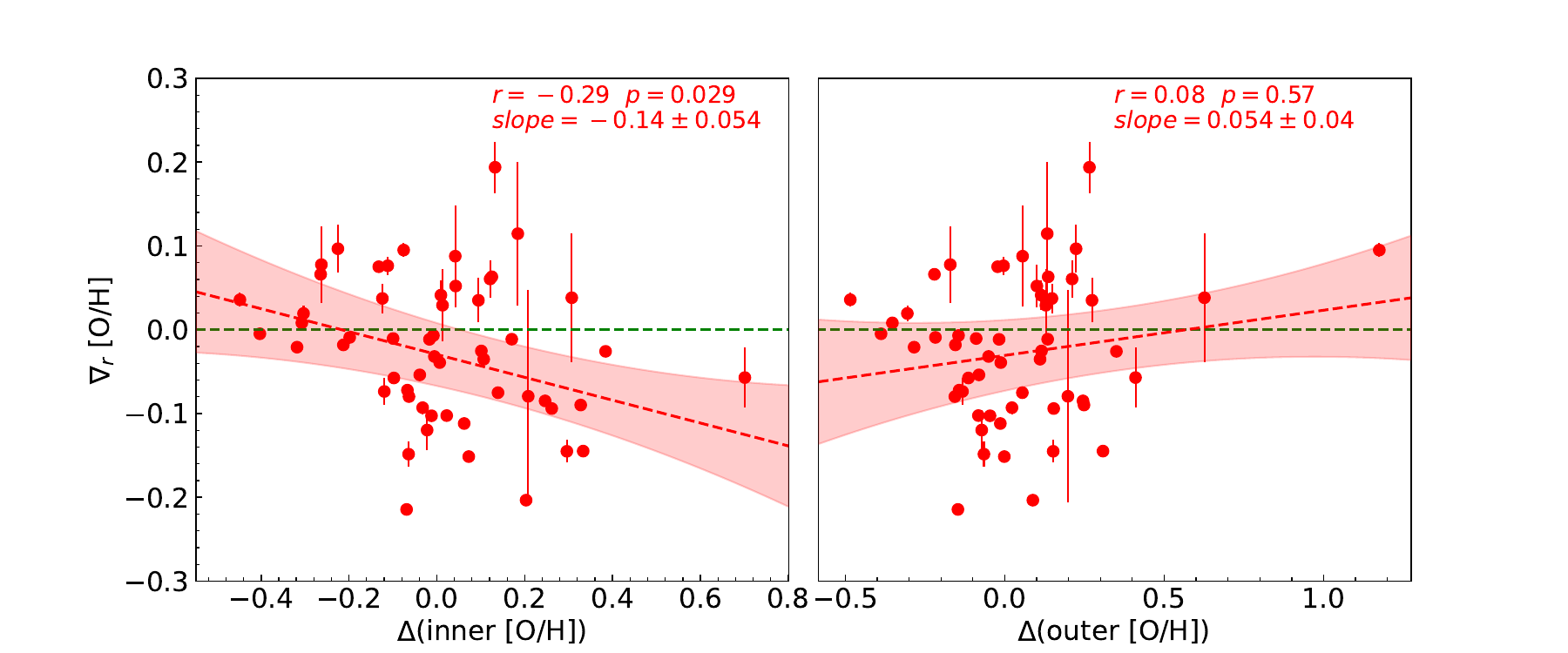}
  \caption{Residual metallicity versus metallicity gradient. Each circle presents one single galaxy in our sample, and the red shaded region represents 1$\sigma$ uncertainties of the best-fit relation, with the slope shown in red dashed line. In both panels, the X-axis represents the metallicity residual of the inner region or outer region of the galaxy in the left or right panel, respectively. The results of best linear fit are shown in the upper right corner in each panel.}
  \label{fig:delta}
\end{figure*}

The variation of metallicity gradient with stellar mass may be primarily driven either by a more significant drop of metallicity at smaller galactocentric radii or a relative enhancement of metallicity at larger radii. To explore this issue, we divide each galaxy into inner and outer regions using the 1/2 $R_{\rm e}$ boundary and sum up the emission line flux in these regions separately to calculate their metallicities. Given the general correlation between gas-phase metallicity and galaxy stellar mass (MZR), we remove a best-fit linear dependence of global metallicity on stellar mass for our galaxies and explore the correlation between residual metallicity and the metallicity gradient.\par

Fig. \ref{fig:delta} illustrates the relationship between residual metallicity and metallicity gradient for the inner and outer galaxy regions separately. A clear negative correlation exists between the metallicity gradient and residual metallicity for the inner regions, with a Spearman's correlation coefficient of $r \simeq -0.29$. No correlation is observed between the residual metallicity of the outer galaxy regions and the metallicity gradient. This result suggests that the flattening of metallicity gradient is primarily caused by a more significant drop of metallicity toward smaller galactic radii, rather than a relative enhancement of metallicity at larger radii.

\subsection{Dependence of the metallicity gradient on secondary parameters} \label{sec:second}
The spatial distribution of metallicities is in principle regulated by several factors, such as the in-situ star formation (metal generation), turbulence driven metal mixing, and metallicity dilution induced by metal-poor gas inflow or metal-enriched gas outflow. As will become clear later, metallicity gradients have the strongest correlation with stellar mass. However, it is not clear what is the physical driver of metallicity gradients in galaxies of different masses and what drives the substantial scatter of metallicity gradients at given stellar mass. In this section, we further explore the connection between metallicity gradients and other galaxy properties. To make a fair comparison of galaxies with different mass and size, we focus on exploring the metallicity gradients normalized by the effective radius $R_{e}$. However, unless otherwise explicitly noted, the general conclusion in this paper is not affected by the way metallicity gradient is expressed.

\subsubsection{Dependence on the gaseous velocity field}
\label{sec:vel}
\begin{figure*}
  \centering
  \includegraphics[width=\textwidth]{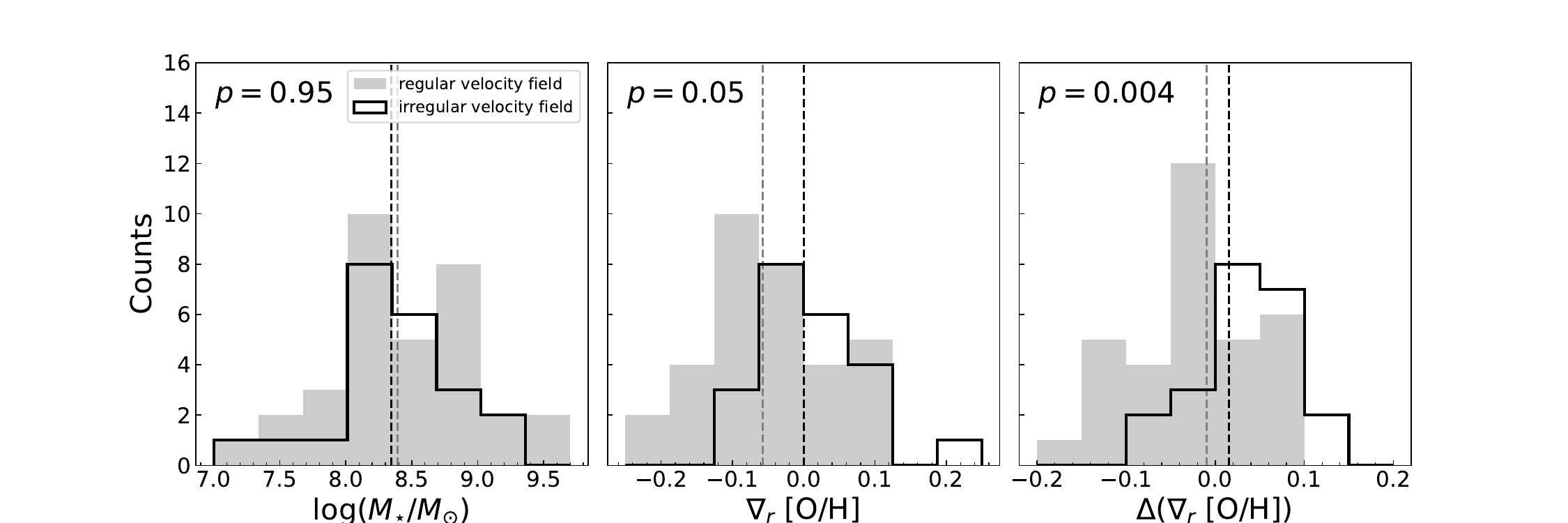}
  \caption{Histogram of metallicity gradient distribution of galaxies with or without regular velocity field. The right panel is for the distribution of residual metallicity gradient after removing the best-fit linear dependence on stellar mass. The gray filled histograms represent galaxies with a clear rotation velocity field, while the black open histograms represent galaxies without a regular gaseous rotational velocity field. The two vertical dotted vertical lines in each panel represent the median values of the two subsamples, and the values on the upper left corner correspond to the p-value from the Kolmogorov-Smirnov tests conducted between galaxies with or without regular velocity field.}
  \label{fig:rot}
\end{figure*}

\begin{figure*}
  \centering
  \includegraphics[width=\textwidth]{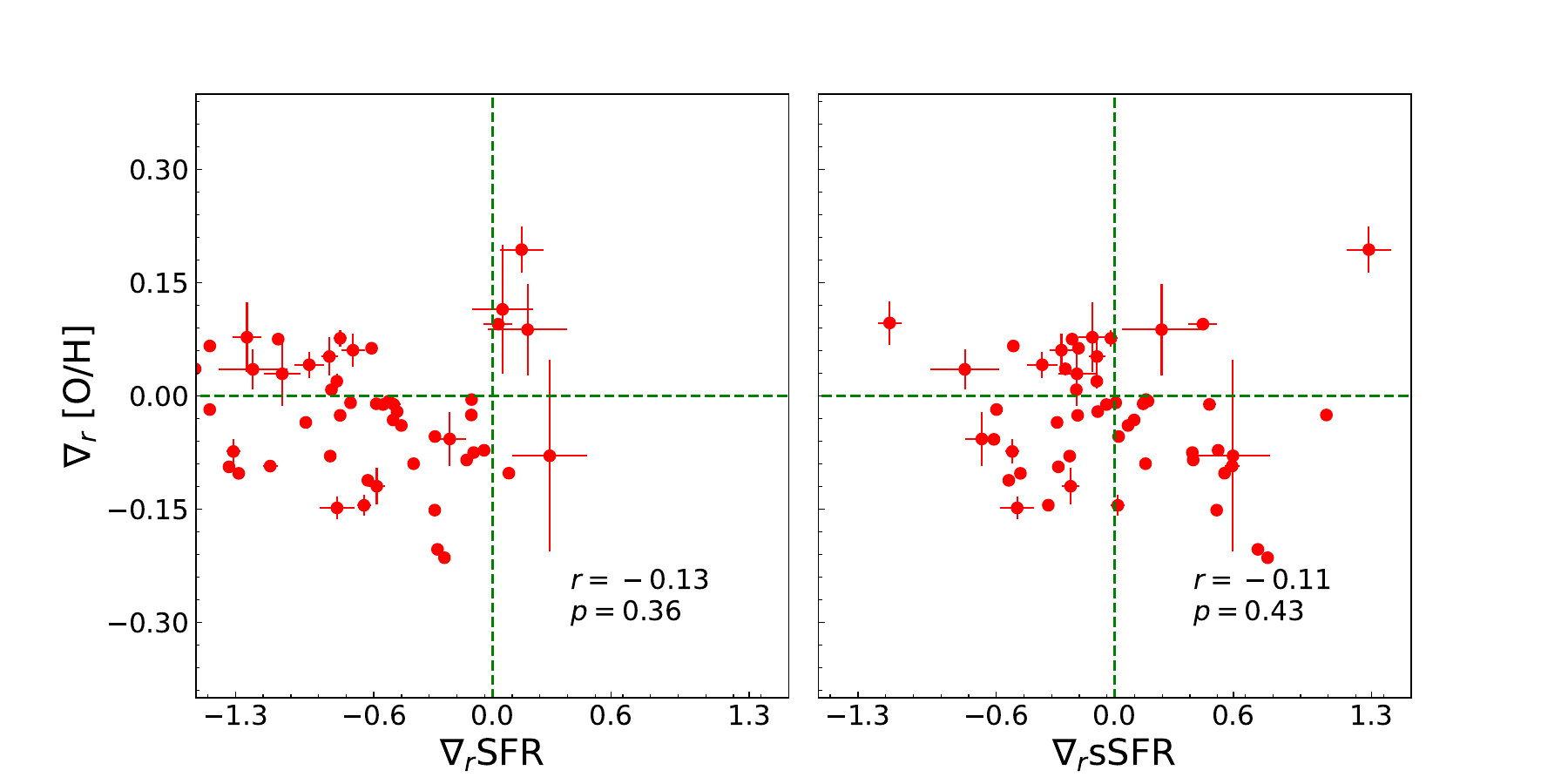}
  \caption{Metallicity gradient versus SFR/sSFR surface density gradients of our sample galaxies. The circles represent the galaxies measured in this work. The green dashed line represents when the gradient is 0. The Spearman correlation test is also given at the right bottom corner in each panel.}
  \label{fig:sfrg}
\end{figure*}

\begin{figure}
    \centering
    \includegraphics[width=\columnwidth]{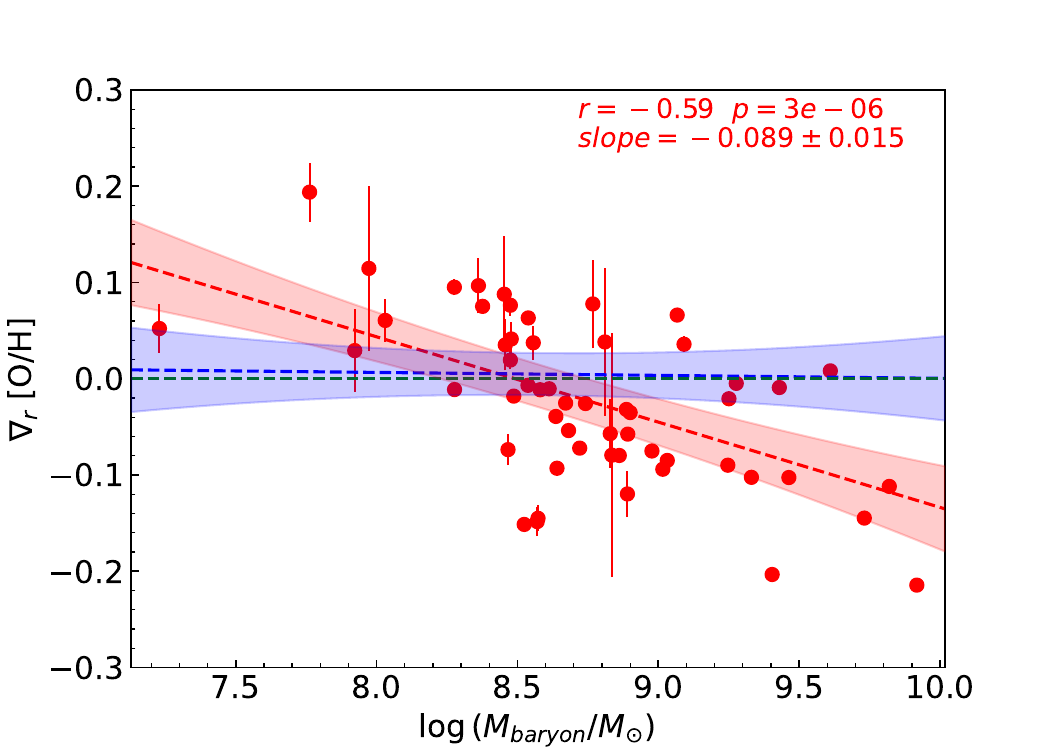}
    \caption{Metallicity gradient as a function of the total baryonic mass. The red circles represent the galaxies measured in this work. The red dashed line and shaded region represent the best-fit linear relation and its 1-$\sigma$ uncertainties. The blue dashed line represents the best-fit linear relation between residual metallicity gradient (after removing the stellar mass dependence) and baryonic mass, and the blue shaded region represents 1$\sigma$ uncertainties of the best-fit relation, with the slope shown in blue. The green horizontal dashed line marks a metallicity gradient of 0.}
    \label{fig:baryon}
\end{figure}

We first classify our galaxies into two subsamples according to the regularity of H$\alpha$ velocity field. A regular velocity field means an overall velocity gradient along the photometric major axis across the main body of galaxies, which is presumably driven by disk rotation, whereas an irregular gaseous velocity field is most likely due to significant disturbance of the interstellar medium by either outflow or inflow activities. Our classification is based on a visual inspection of the H$\alpha$ velocity field and the velocity variation along the major and minor axes, as exemplified in Fig. \ref{fig:metal_measure}, which suffices for our purpose. It turns out 33 galaxies exhibit regular velocity field and the remaining 22 galaxies exhibit irregular velocity field. We emphasize that the classification of regular and irregular gaseous velocity field serves as an indicator of the dynamical status of the interstellar medium (ISM), rather than the presence or lack of a galaxy disk. 

In the left panel of Fig. \ref{fig:rot}, we show the stellar mass distributions of the two subsamples. It is clear that galaxies with regular velocity field have a stellar mass range and distribution similar to those with irregular velocity field in our sample. In the middle panel, we compare the metallicity gradient distribution of the two subsamples with or without regular velocity field. Since metallicity gradients are correlated with galaxy stellar mass, we also explore the metallicity gradient difference of the two subsamples after removing the best-fit linear stellar mass dependence of the metallicity gradient.

The subsample with irregular velocity field has a zero median metallicity gradient, and the subsample with regular velocity field has a median gradient of $-0.057$ dex $R_e^{-1}$. After controlling for stellar mass, the two subsamples still show systematically different distributions of metallicity gradients. Particularly, the galaxies with regular velocity field have a median $\Delta(\nabla_{R_e})$ of $-$0.009 dex $R_e^{-1}$, whereas it is 0.015 dex $ R_e^{-1}$ for those with irregular gaseous velocity field. The Kolmogorov–Smirnov test suggests that the two subsamples are not likely to be drawn from the same underlying distribution (p-value = 0.004, $D$ = 0.348). Nevertheless, we note that the mass-metallicity correlation shown in Fig. \ref{fig:mzgr} remains virtually unchanged when galaxies with irregular gaseous velocity fields are excluded. \par

\cite{Ma2017} invokes a toy model to explore the connection between metallicity gradient and disk regularity, which is updated by \cite{sun2024}. In their model, if the metals do not mix efficiently between radial annuli, the gas-phase metallicity follows the mass fraction of stars, resulting in a negative metallicity gradient. Galaxies with irregular gaseous velocity field may be strongly perturbed by violent processes, such as mergers, rapid gas inflows, and strong feedback-driven outflows, which may substantially disturb pre-existing rotation-dominated velocity field and cause efficient gas re-distribution on galactic scales and thus leads to non-negative metallicity gradients. Galaxies with regular rotational velocity field but flat metallicity gradients may be in a transition stage, e.g. during a significant gas inflow after which a negative metallicity gradient will build up at a later time.\par 

For our sample of ordinary dwarf galaxies in the local universe, galaxy mergers and violent gas accretion are unlikely to play a significant role (see also Sect. \ref{subsec:environment}). Instead, the irregular gaseous velocity field is most likely a result of stellar feedback disturbing the interstellar medium. Furthermore, the fact that the least massive galaxies in our sample predominantly exhibit positive metallicity gradients suggests that the above-mentioned ``transitional'' stage (if any) has a prolonged duty cycle for these galaxies.

\subsubsection{Dependence on radial gradient of current star formation}
\label{sec:sfrg}

\begin{figure*}
  \centering
  \includegraphics[width=\textwidth]{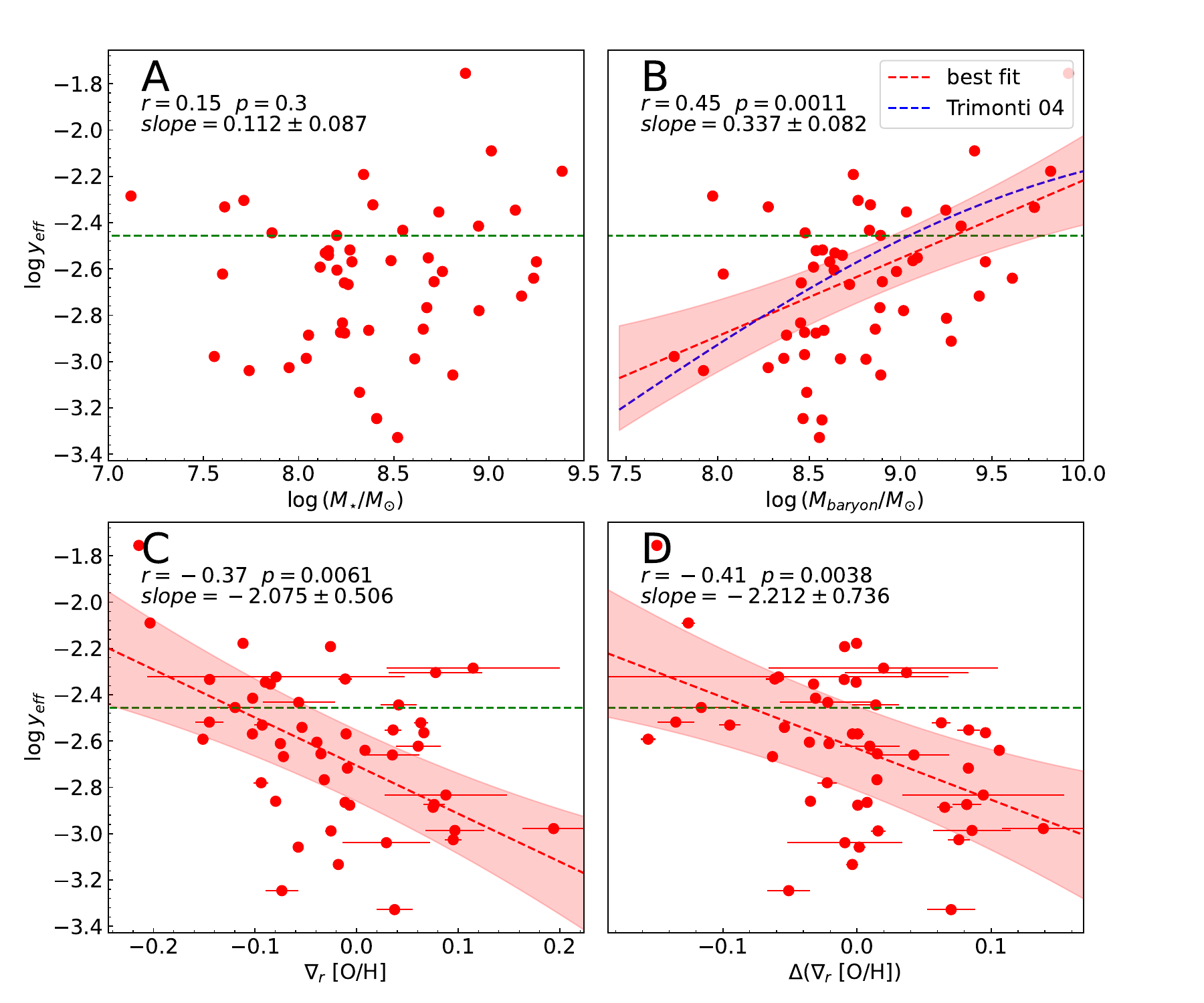}
  \caption{Panels A, B, C, D (indicated in the upper left corner of each panel) respectively show the relation between effective yield and stellar mass, total baryonic mass (stellar+neutral atomic mass), metallicity gradient and residual metallicity gradient (after removing the stellar mass dependence). In each panel, the green dashed horizontal line represents the true stellar oxygen yield derived by \cite{Pilyugin2007}, and the red dashed line and the shaded region are the best-fit linear relation and its 1-$\sigma$ uncertainties. The Spearman's rank correlation coefficient $r$ and p-values and the best-fit slope are given in the upper left corner of each panel. In panel B we also plot the best fit relation from \protect\cite{tremonti2004}.}
  \label{fig:yeff}
\end{figure*}

According to the classical "inside-out" disk growth paradigm \citep[e.g.,][]{Chiappini2001}, gas accumulation and consumption (through star formation) are faster at smaller galactocentric radii, which would naturally predict a negative metallicity gradient if there is negligible radial migration of matter. In this paradigm, the star formation rate profile has a direct consequence on the metallicity gradient \citep[e.g.,][]{Pilkington2012}. In this section, we explore the connection of the radial gradients of metallicity and star formation.

Figure \ref{fig:sfrg} shows the metallicity gradients as a function of the radial gradients of SFR surface density (left panel) and sSFR (right panel). The correlation with SFR surface density and sSFR radial gradients is very weak, with a Spearman's rank correlation coefficient $r$ of $-$0.13 and $-$0.11, and a p-value of 0.36 and 0.43, respectively. This strongly suggests that physical processes other than in-situ star formation, such as metal mixing/migration or radially differential dilution of metallicity, play more important roles in shaping the metallicity gradients. The lack of correlation rules out the possibility that metal-poor gas inflow to galactic center drives the flat or positive metallicity gradient, as otherwise we would expect enhanced central star formation and thus steeper (i.e., more negative) radial gradient of SFR surface density in galaxies with flatter or more positive metallicity gradient.

\subsubsection{Dependence on the baryonic mass} 
\label{sec:dep_baryon}
\begin{figure*}
    \centering
    \includegraphics[width=1\textwidth]{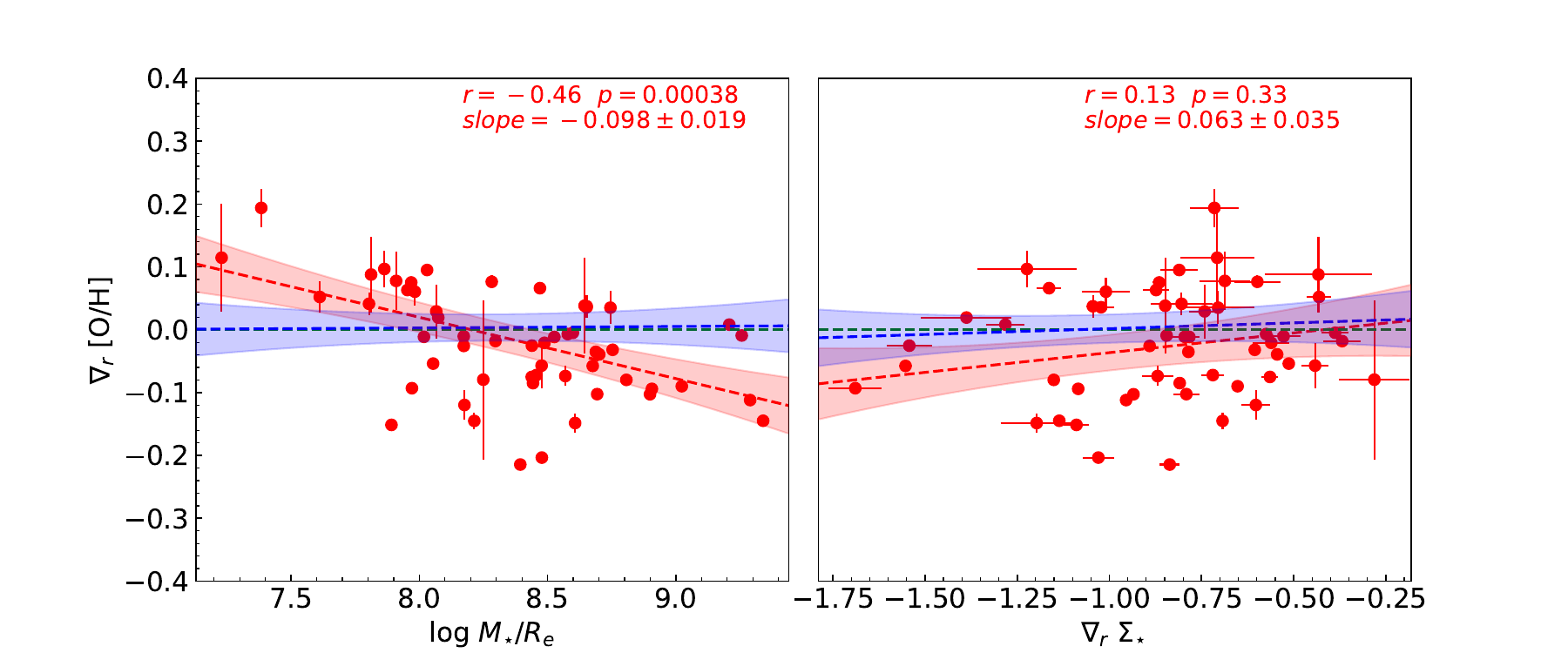}
    \caption{The dependence of metallicity gradient on stellar gravitational potential, as represented by $M_{\star}/R_e$, and radial gradient of the stellar mass surface density. The meaning of the different lines and colors are similar to that in Fig. \ref{fig:baryon}. The apparent correlation of metallicity gradients with $M_{\star}/R_e$ or stellar mass surface density disappears once the dependence on stellar mass is removed.}
    \label{fig:potential}
\end{figure*}

We look into the correlation between the metallicity gradient and the total baryonic mass of our galaxies. Here the baryonic mass includes the stellar mass and cold interstellar gas mass (based on HI 21 cm emission line and star formation rate). We intend to explore which of the two (baryonic mass and stellar mass) has the strongest correlation with metallicity gradient.\par

We plot the result in Fig. \ref{fig:baryon}. In this figure, we can see a negative correlation between the baryonic mass and the metallicity gradient. The slope of the best-fit linear relation is $-$0.089$\pm$0.017, slightly lower than that between stellar mass and metallicity gradient in unit of dex $R_{\rm e}^{-1}$. The Spearman's rank correlation coefficient $r$ is $-$0.58 with a sigma of 0.073. Although the correlation coefficient here is slightly higher than that with stellar mass, the difference is within the 1 $\sigma$ uncertainties. 
Furthermore, we also explore the residual metallicity gradient correlation with baryonic mass by removing the best-fit linear dependence of gradient on stellar mass (Sect. \ref{sec:mzgr}). Specifically, for each galaxy, we derive the deviation $\Delta(\nabla_{R_e})$ of its metallicity gradient from the best-fit relation and perform a linear fitting of the residual gradient vs. baryonic mass. The best-fit linear relation is over-plotted (in blue color) in Fig. \ref{fig:baryon}.
We find that the Spearman's rank correlation coefficient $r$ is $-$0.046, and the best-fit slope is 0.011$\pm$0.022. So there is virtually no correlation between metallicity gradient and baryonic mass, once the stellar mass dependence is removed.

To further examine the lack of intrinsic correlation between baryonic mass and metallicity gradients, we calculate the partial correlation coefficient between metallicity gradients and gas mass while controlling for stellar mass. The resulting $r$ is $-$0.093, indicating no intrinsic correlation between gas mass and metallicity gradients. Above all, stellar mass is the driver of the apparent correlation between metallicity gradients and baryonic mass.

\subsubsection{Dependence on effective yield}

As mentioned in Sect. \ref{sec:eff_measure}, the effective yield $\rm y_{eff}$ serves as a diagnostic parameter for probing outflows or gas inflows. Here we explore the connection between $\rm y_{eff}$ and metallicity gradients, in an attempt to probe the relevance of inflow and outflow. While $\rm y_{eff}$ has been routinely estimated in the literature, we emphasize that the fundamental assumption of instantaneous and homogeneous chemical mixing for deriving $\rm y_{eff}$ is often violated in reality, as evidenced by the metallicity gradients and differential spatial distributions of gas and stars commonly observed in galaxies. Nevertheless, a galaxy operates as an interconnected ecosystem, where gas inflows and outflows facilitate exchange between the inner and outer regions. Therefore, a global $\rm y_{eff}$ still provides helpful insights into the overall baryon cycling process.

In Fig. \ref{fig:yeff}, the upper two panels show the distribution of our galaxies on $\rm y_{eff}$ vs. stellar mass and $\rm y_{eff}$ vs. baryonic mass planes. The true stellar yield derived by \cite{Pilyugin2007} is over-plotted as a reference (horizontal green dashed lines in Fig. \ref{fig:yeff}). The best-fit relation between $\rm y_{eff}$ and baryonic mass from \cite{tremonti2004} is also overplotted for comparison. Most of our galaxies (40 in 55) have lower $\rm y_{eff}$ than the true stellar yield, which hints at a significant influence by gas outflow or inflow. $\rm y_{eff}$ has no correlation with stellar mass ($r$ $\simeq$ 0.1) but has a moderate correlation with the baryonic mass ($r$ $\simeq$ 0.4). A correlation between $\rm y_{eff}$ and baryonic mass has already been found by \cite{tremonti2004}. If assuming baryonic mass is a better tracer of the total galaxy mass than stellar mass \citep[as evidenced by the existence of a tight baryonic Tully$-$Fisher relation;][]{mcgaugh2012}, the correlation may imply a connection between gravitational potential well and $\rm y_{eff}$, and thus may support the scenario of metal-enriched outflow driving lower $\rm y_{eff}$, because shallower potential well in lower mass galaxies favors a stronger stellar feedback driven outflow. It is however intriguing that \cite{vanZee2006} did not find a correlation of $\rm y_{eff}$ with dynamical mass for their sample of (mainly) low mass galaxies.

\begin{figure*}
  \centering
  \includegraphics[width=\textwidth]{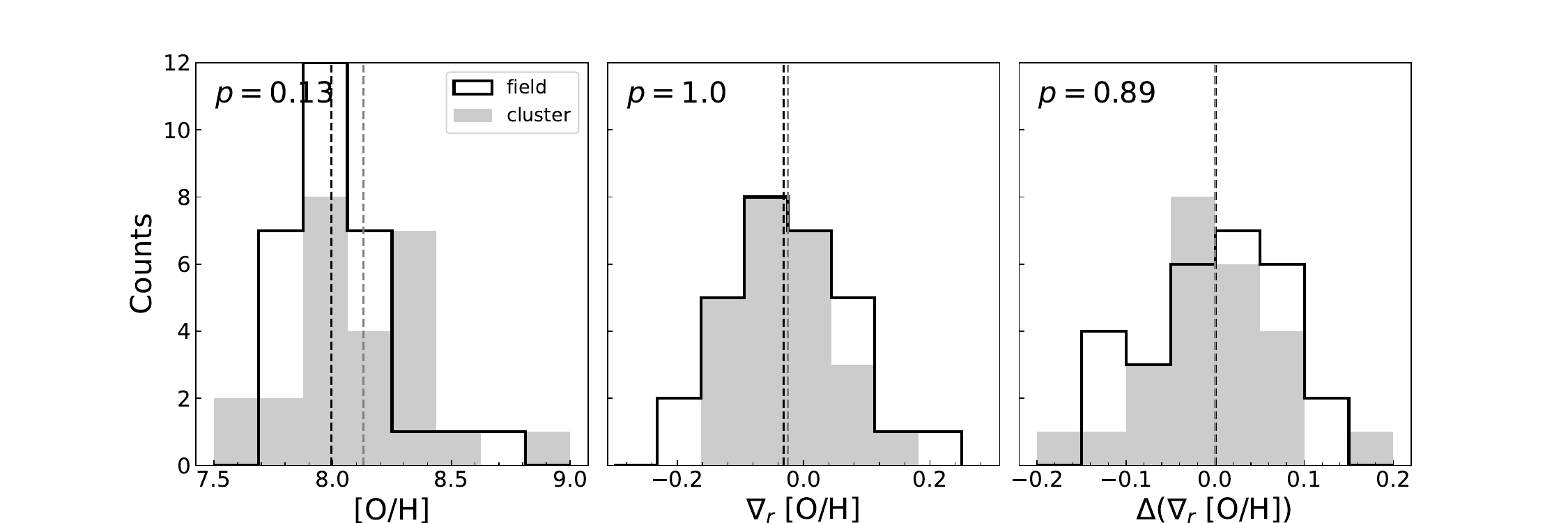}
  \caption{Environmental dependence of the metallicity, metallicity gradient and metallicity gradient residual (by removing the best-fit linear relation between gradient and stellar mass) distributions. The gray shaded histograms represent the cluster subsample, and the open histograms represent the field subsample. The vertical black (gray) dashed line marks the median values of field (cluster) subsample. The values on the upper left corner correspond to the p-value from the Kolmogorov-Smirnov tests conducted between galaxies in different environments.}
  \label{fig:environment}
\end{figure*}

\begin{figure}
  \centering
  \includegraphics[width=\columnwidth]{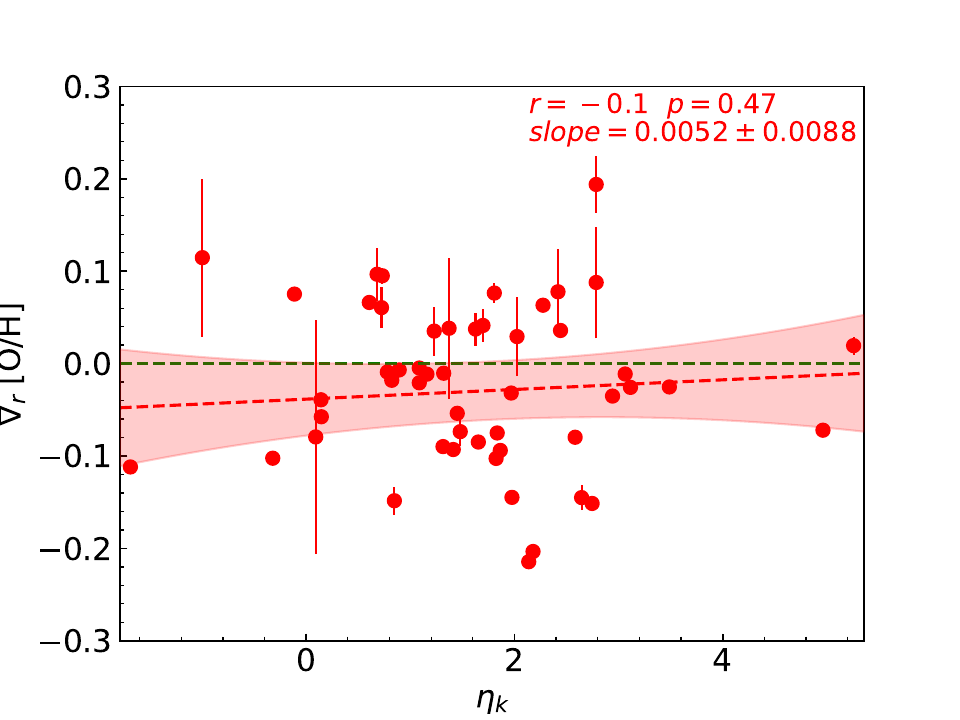}
  \caption{The metallicity gradients as a function of the projected local number density of galaxies. Circles represent individual galaxies, and the red shaded region represents 1$\sigma$ uncertainties of the best-fit relation, with the slope shown in red dashed line. The Spearman correlation coefficient and best-fit slope are given in the upper right corner.}
  \label{fig:pden}
\end{figure}

The lower two panels of Fig. \ref{fig:yeff} show the relation between $\rm y_{eff}$ and metallicity gradients of our galaxies. Particularly, in the lower right panel, the stellar mass dependence of metallicity gradient has been removed as in previous section. As indicated in the lower panels, the metallicity gradient shows a moderate ($r$ $\simeq$ 0.4) and significant (p-values $<$ 0.01) negative correlation with $\rm y_{eff}$. It appears that the metallicity gradient shows a slightly stronger correlation with $\rm y_{eff}$ once controlling for galaxy stellar mass. Therefore, $\rm y_{eff}$ partially explains the scatter of the galaxy stellar mass$-$metallicity gradient correlation. We will revisit this point later in the paper.

\subsubsection{Dependence on stellar gravitational potential}

$M_{\star}/R_e$ has been often used as a proxy for galaxy stellar potential well in the literature \citep[e.g.,][]{Sanchez-Menguiano2024}. Here we explore the relation between $M_{\star}/R_e$ and metallicity gradients of our galaxies. In addition, we also explore the connection with the radial slope of the stellar mass surface density profile.
The results are shown in Fig. \ref{fig:potential}. The metallicity gradient has a moderate negative correlation with $\log(M_{\star}/R_{\rm e})$, but no correlation with the stellar mass surface density gradient. Nevertheless, once controlling for the galaxy stellar mass dependence, the correlation with $\log(M_{\star}/R_{\rm e})$ disappears.

\subsubsection{Dependence on environment}
\label{subsec:environment}
\begin{figure}
    \centering
    \includegraphics[width=\linewidth]{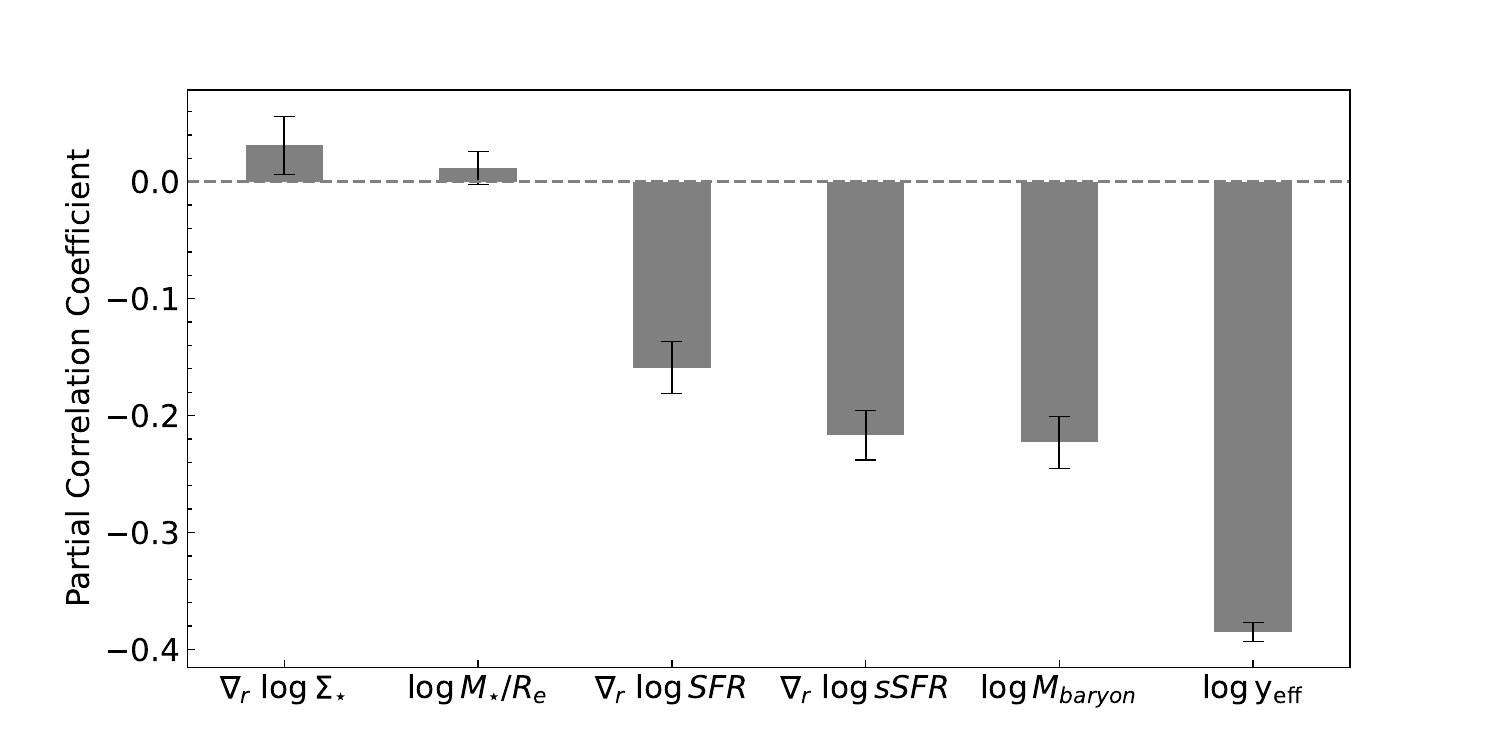}
    \caption{Partial Correlation Coefficients between metallicity gradients and various parameters by controlling for galaxy stellar mass, $M_{\star}$. Filled bars show the Spearman's rank correlation coefficient, and the uncertainties are computed via bootstrap random sampling.}
    \label{fig:pcc}
\end{figure}

\begin{figure}
    \centering
    \includegraphics[width=\linewidth]{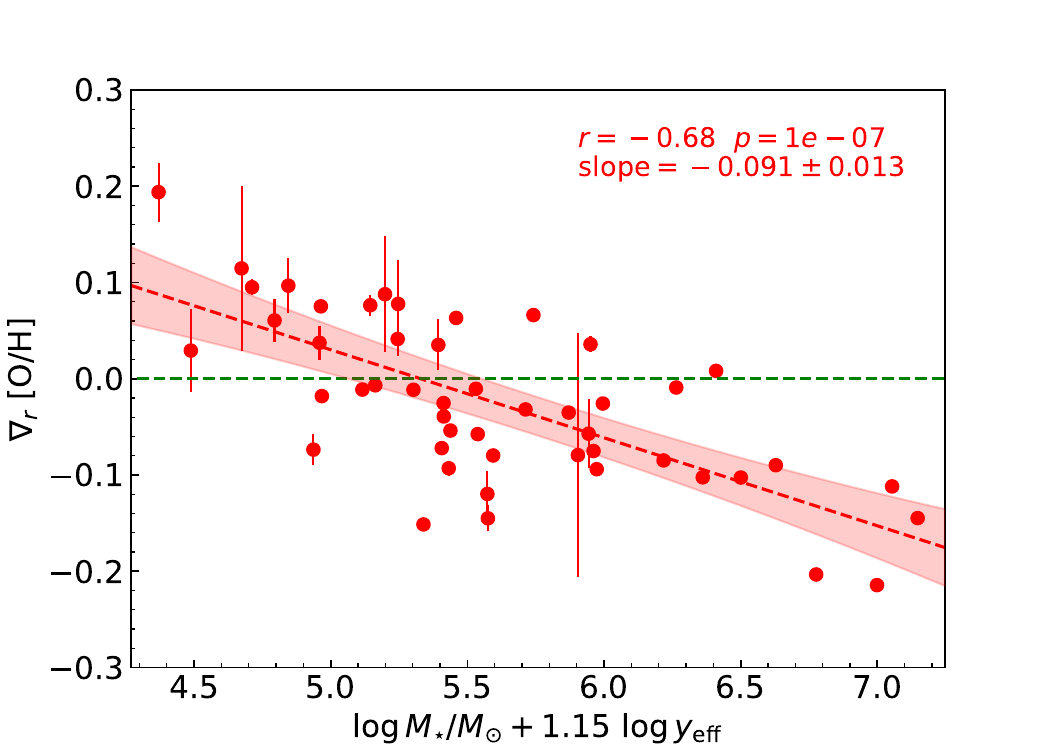}
    \caption{Relation between metallicity gradients and $\mu_{1.15} = \log {M_{\star}}/{M_{\odot}}+1.15 \times \log \rm y_{eff}$. The coefficient of $\log \rm y_{eff}$, $\alpha$ = 1.15, minimizes the scatter of metallicity gradients at given $\mu_{1.15}$. The best-fit slope, Spearman's rank correlation $r$ and p-value are given at the upper right corner of this figure. The red dashed line and shaded region represent the best-fit linear relation and its 1-$\sigma$ uncertainties. The combination parameter $\mu_{1.15}$ exhibits a {\em strong} correlation with metallicity gradients, representing the tightest correlation observed with metallicity gradients. See Sect. \ref{sec:pcc_rfr} for details.}
    \label{fig:mzgefr}
\end{figure}

Both local and large scale environment may affect the metallicity distribution in galaxies. For instance, ram pressure or tidal disruption/compression in group and cluster environment may cut off cosmic gas accretion, change the spatial distribution of gas inside galaxies and thus affect the star formation distribution and metallicity distribution \citep[e.g.,][]{lara2022}; There is also evidence of disturbed gaseous disk in close pairs of galaxies, suggesting the influence of local environment on metallicity distribution. In this subsection, we explore the environmental dependence of the metallicity gradients. Specifically, for the large scale environment, we divide the whole sample into two environmental types: field galaxies (26) and cluster galaxies (29), where galaxies located in groups and clusters are both referred as cluster galaxies for brevity. And for the local environment, following \cite{argudo2015}, we define the projected galaxy number density parameter as:

\begin{equation}
\label{eq:pden}
        \eta_{k} \equiv \log \left(\frac{k-1}{\operatorname{Vol}\left(d_{k}\right)}\right)=\log \left(\frac{3(k-1)}{4 \pi d_{k}^{3}}\right)
\end{equation}

where $\eta_{k}$ is the projected physical distance to the $k$th nearest neighbor. Here we adopt 5 as the value of $k$. The farther the $k$th nearest neighbor, the lower the projected number density $\eta_{k}$.\par

The comparison of metallicity, metallicity gradient and metallicity gradient residual distributions of field and cluster galaxies is shown in Fig. \ref{fig:environment}. The open histograms in Fig. \ref{fig:environment} represent the distributions of field galaxies, while the gray filled histograms represent the cluster galaxies. From the metallicity distributions (left panel), we see a median metallicity difference between the field and cluster galaxies, which is probably attributed to the inhomogeneous nature of the current sample. In addition, compared to the field galaxies, the median metallicity gradient (middle panel) of cluster galaxies is slightly more negative than field galaxies. In order to alleviate the effect of mismatch between the mass/metallicity distributions of our field and cluster subsamples, we subtract the best-fit stellar mass$-$metallicity gradient linear relation from the measured metallicity gradient, and present the metallicity gradient residual distribution in the right panel of Fig. \ref{fig:environment}. The field and cluster galaxies have virtually the same median metallicity gradient residual ($\sim$ $-$0.01). To further quantify the difference or similarity of the gradient residual distributions of the two subsamples, we run a Kolmogorov–Smirnov test for a null hypothesis that the two being drawn from the same underlying distribution, and find a p-value of 0.89, which means that the two subsamples are consistent with being drawn from the same underlying distribution. Therefore, the large-scale environment appears not relevant in shaping the metallicity gradient of these galaxies.\par

In Fig. \ref{fig:pden}, we plot the metallicity gradients as a function of projected galaxy number density. We find no significant correlation between metallicity gradient and projected galaxy number density, with a Spearman's rank correlation coefficient $r$ $=$ $-0.065$ and p-value $=$ 0.65. This implies that the local environment also does not affect the metallicity gradient of our galaxies. \par

\subsection{Partial correlation analysis and a combination of stellar mass and \texorpdfstring{$\rm y_{eff}$}{}}\label{sec:pcc_rfr}

In the previous sub-sections, we have explored the connection between metallicity gradients and various galaxy parameters. Here we summarize the strength of these connections by presenting the Spearman partial correlation coefficients (PCC; controlling for galaxy stellar mass) in Fig. \ref{fig:pcc}. The galaxy parameters that we have considered include the effective yield, $\rm y_{eff}$, baryonic mass, logarithmic stellar gravitational potential $\log$($M_{\star}/R_e$), radial gradients of stellar mass surface densities, star formation surface densities and sSFR. In Fig. \ref{fig:pcc}, the error bars of the PCC are uncertainties obtained by bootstrap sampling of our galaxy sample.

As can be seen, after controlling for galaxy stellar mass, $\rm y_{eff}$ is the most important and dominant galaxy parameter in predicting the metallicity gradient of a galaxy, and the second most relevant parameter is the baryonic mass. The PCC analysis indicates that the metallicity gradients have a moderate correlation ($r$ $\simeq$ 0.4) with $\rm y_{eff}$, whereas the other secondary parameters have much weaker correlation ($r$ $\lesssim$ 0.2).\par

Now that we know that effective yield is the second most relevant parameter that correlate with metallicity gradient, we perform a linear least-squares fitting of metallicity gradient as a function of a combination of logarithmic stellar mass and effective yield, that is,
\begin{equation}
\label{eq:mzgef}
\nabla_{r}[\mathrm{O} / \mathrm{H}] = \alpha \times f(M_{\star},\rm y_{eff})+\beta,
\end{equation}
where $f(M_{\star},\rm y_{eff})$ is a linear combination of logarithmic stellar mass and effective yield, defined as \begin{equation}
\label{eq:fmyeff}
f(M_{\star},\rm y_{eff}) = \log \left(\frac{M_{\star}}{M_{\odot}}\right)+\gamma \times \log \left(\rm y_{eff}\right)
\end{equation}

To search for the optimal parameter values of $\alpha$, $\beta$, $\gamma$, we traverse the $\gamma$ values (accurate to two decimal places), and for each $\gamma$, perform a linear least-squares fitting to our galaxies and find $\alpha$ and $\beta$ values that minimize the scatter of $\nabla_{r}[\mathrm{O} / \mathrm{H}]$ around the best-fit linear relation (Eq. \ref{eq:mzgef}). From this iterative linear least-squares fitting, we find the optimal parameter values of $\alpha$, $\beta$ and $\gamma$ are $-$0.091$\pm$0.013, 0.45$\pm$0.17, and 1.15. 

We note that the above best-fit $\alpha$ coefficient value is equal to that of the linear relation of $\nabla_{r}[\mathrm{O} / \mathrm{H}]$ vs. $\log {M_{\star}}/{M_{\odot}}$ (Sect. \ref{sec:mzgr}), but the Spearman's rank correlation coefficient with $\nabla_{r}[\mathrm{O} / \mathrm{H}]$ increases from $\sim$ $-0.58$ to $-0.67$. The scatter of $\nabla_{r}[\mathrm{O} / \mathrm{H}]$ around the best-fit relations is reduced from 0.068 to 0.060 after involving $\log \rm y_{eff}$. The mean measurement uncertainty of metallicity gradient is 0.006 dex $R_{e}^{-1}$. Therefore, $\log \rm y_{eff}$ accounts for 22\% (1$-(0.060^{2}-0.006^{2})/(0.068^{2}-0.006^{2})$) of the scatter of $\nabla_{r}[\mathrm{O} / \mathrm{H}]$$-$stellar mass relation. Fig. \ref{fig:mzgefr} shows the distribution of our galaxies on the 
$\nabla_{r}[\mathrm{O} / \mathrm{H}]$ vs. ($\log {M_{\star}}/{M_{\odot}}+1.15 \times \log \rm y_{eff}$) plane.

\section{Discussion}
\label{sec:discussion}
With the connection between metallicity radial profiles and other galaxy properties explored in the previous section, we here compare our results with previous studies and discuss the possible implications of these results on dwarf galaxy evolution.

\subsection{MZGR at the low mass end}
According to previous studies \citep[e.g.,][]{zartsky1994, wang2022, wang2024}, the well-known exponential disks and negative radial gradients of gas-phase metallicity in massive disk galaxies are the result of inside-out star formation or gradual gas-flow along the disk from outside in. The former scenario suggests that the presence of the exponential disk is a natural result of galaxy star formation regulated by gas distribution in galaxies, with the inner high gas density regions forming stars earlier and more efficiently than the outer disks. A negative metallicity gradient follows naturally in this inside-out disk formation process. The gas-flow scenario attributes the negative metallicity gradients to a gradual metal-enrichment of gas that flows from outside in. These two scenarios are not necessarily mutually exclusive, and both may happen in real galaxies.\par

Physical drivers of metallicity gradient can be constrained by exploring its connection with various galaxy properties. Data from several large IFU surveys have been used to identify Mass$-$Metallicity Gradient Relation (MZGR) in a number of studies (MaNGA, \citep[e.g.,][]{belfiore2017}; SAMI, \citep[e.g.,][]{poetrodjojo2021}). These studies found that, starting from the low galaxy stellar mass end of their samples (log($M_{\star}/M_{\odot}) \sim 9$), the metallicity gradient gradually becomes more negative, with the trend showing a mild curvature around log($M_{\star}/M_{\odot}) \sim 10-10.5$. At the higher mass end (log($M_{\star}/M_{\odot}) > 11$), the metallicity gradients flatten again. We note that the metallicity gradient variation with galaxy mass is found to be generally weak, with no mass dependence found in some studies \citep[e.g.,][]{Lian2018}. For the higher mass end, general view is that the flattening of the metallicity gradient can be ascribed to the general behavior of evolved systems to reach an equilibrium abundance at late times. In this scenario, as galaxy grows, the stellar mass increases, and the flattening should occur ‘inside-out’, with central regions reaching their equilibrium metallicity earlier and having lower gas fraction than do the outer regions \citep[e.g.,][]{belfiore2017, pilyugin2023}. \par

However, for dwarf galaxies (log($M_{\star}/M_{\odot}$) $\lesssim$ 9.5), we found that the situation appears quite different. The metallicity gradients become gradually less negative (i.e., flatter or more positive) toward the lower-mass end of dwarf galaxies. Instead of being attributed to a progressive saturation of metal production with radius as mentioned above for massive galaxies, the finding for dwarf galaxies is more likely due to a more significant metal mixing and transport within lower mass dwarf galaxies. As shown in Sect. \ref{sec:vel}, galaxies with irregular gaseous velocity fields tend to have less negative or more positive metallicity gradients. The irregular gaseous velocity field suggests significant perturbations in the galaxy-scale distribution of the interstellar medium (ISM), potentially driving the redistribution of metals, primarily from the inner regions outward. This is supported by the greater degree of metallicity suppression observed in the central regions compared to the outskirts of our galaxies toward the lower mass end (Fig. \ref{fig:delta}).
We note that radial mixing of metals has also been invoked to explain the shallower metallicity gradients of older stellar populations of the Milky Way-like galaxies \citep[e.g.,][]{graf2024}.\par

\subsection{Stellar feedback as an important factor reshaping metallicity distribution}

Understanding how gas flows and recycles is crucial in studying galaxy evolution. Gas inflow is generally challenging to identify in observations. While gas outflows have been identified in numerous studies, the overall significance of their impact on the baryonic cycling within galaxies is not clear. In this section, we discuss how gas flow influences metallicity gradients, based on the analysis in Sect. \ref{sec:second}.

We first consider the inflow of metal-poor gas. If the infalling gas is relatively metal-poor, it dilutes the gas enriched by in-situ star formation, thereby reduces metallicity, and at the same time may enhance local star formation activities. This may therefore lead to an anti-correlation between metallicity and star formation rate (SFR) if gas inflow dominates. However, as exemplified by the two galaxies NGC1796 and NGC1705 (Sect. \ref{sec:metal_measure}), we do not find such negative correlation between metallicity and SFR in most star-forming regions in our sample, and only few regions in galaxies like NGC1705 (a starburst dwarf; at 1 $R_{\rm e}$ radius) exhibit high SFR and low metallicity. Our sample's scarcity of such anti-correlation between SFR and metallicity indicates that metal-poor gas inflow may not be the dominant factor in dwarf galaxies and contributes little to the metallicity gradients in general.\par

The lack of correlation with SFR/specific SFR (sSFR) surface density gradients (Sect. \ref{sec:sfrg}) again supports our point of view. If we assume that metal-poor gas inflow is the dominant factor (as compared to metal-enrichment outflow) in dwarf galaxies leading to inverted metallicity gradients, we expect steeper SFR/sSFR surface density gradients where the metallicity gradient is flatter. However, as demonstrated in Fig. \ref{fig:sfrg}, we did not find a strong correlation between these two gradients, suggesting that the inflow scenario may not be the dominant factor shaping metallicity distribution in most dwarf galaxies.\par

Besides gas inflow, other processes, such as metal-enriched gas outflow, turbulence-driven spatial mixing, may also affect the metallicity distribution within galaxies. In this regard, one commonly mentioned scenario is the so-called "galactic fountains", where metal-enriched gas outflow launched (by stellar feedback) from smaller galactocentric radii is finally deposited at larger radii. \citep[e.g.,][]{Gibson2013, Ma2017, Tissera2022}. \par

Previous studies \citep[e.g.,][]{tremonti2004, daddi2007, lara2019} suggest that metal-enriched outflow is the primary factor that results in the observed lower effective yield of metals $\rm y_{eff}$ in galaxies of lower baryonic mass. Therefore, $\rm y_{eff}$ may serve as an indicator of outflow efficiency. We find a significant correlation between metallicity gradient and $\rm y_{eff}$, even after controlling for galaxy mass. This suggests that metal-enriched outflow is indeed an important mechanism in re-distributing metals in dwarf galaxies.\par

Regarding outflow, we also explore the correlation between stellar gravitational potential, baryonic mass and metallicity gradient. In principle, feedback-driven outflow is expected to be more efficient in galaxies with shallower gravitational well. While the finding that these parameters ($\log(M_{\star}/R_{\rm e})$, $M_{\rm baryon}$) have a significant correlation ($r$ $\sim$ $-0.4 $--$ -0.6$) with metallicity gradient may suggests a non-negligible role the gravitational potential well in (re)shaping metallicity gradient. it is however intriguing that the correlations disappear after controlling for galaxy stellar mass. This indicates a non-trivial connection between stellar feedback and metallicity distribution. The effective yield correlates with metallicity gradient, but it only accounts for 22\% of the scatter of MZGR. We speculate that outflow is probably just one way that feedback affects metallicity distribution. Other metal mixing and transport process, such as feedback-fed turbulence of ISM, may also play important roles. 

Some studies attributed lower oxygen abundance at the centers of some dwarf galaxies to pristine gas infall rather than metal-enriched outflow \citep[e.g.,][]{kewley2010, chung2023}. This may be in line with our finding of a greater degree of suppression of metallicities toward smaller galactic radii of lower mass galaxies (Fig. \ref{fig:delta}). Gas inflow may be important in some dwarf galaxies, especially the starburst ones \citep[e.g.,][]{tang2022}, but it may not be a dominant factor that determines gas metallicity gradient of dwarf galaxies in general.\par

\subsection{The relevance of environment}
Environment can have a significant impact on galaxy evolution. Galaxies in clusters tend to have higher metallicities than field galaxies \citep[e.g.,][]{ellison2009, lian2019}. However, studies of the effect of environment on gas metallicity gradients have so far been rather limited. While some studies find the metallicity gradient is independent with the density of the environment \citep[e.g.,][]{sanchez2018}, others find galaxies in clusters tend to have flatter metallicity gradients than field galaxies \citep[e.g.,][]{kewley2010, lara2022, franchetto2021}.\par

In this work, we find that field galaxies have a significant lower metallicity than cluster galaxies, consistent with previous results. However, their metallicity gradients exhibit negligible difference when controlling the stellar mass. Moreover, we find no significant correlation between metallicity gradient and the projected local galaxy number density. These results may imply that, gas accretion from intergalactic medium (IGM) or circumgalactic medium (CGM), if any,  may influence overall metallicities, but it is not directly connected to the metallicity distribution within dwarf galaxies. Due to their shallow potential wells, dwarf galaxies are expected to be advection-dominated rather than governed by cosmological accretion \citep[][]{sharda2021}. This means that internal metal redistribution processes are more important than external gas accretion in shaping the radial profile of metallicity in dwarf galaxies.

\section{Summary}
\label{sec:summary}
In this paper, we present a study of the radial gradients of gaseous metallicities of a sample of nearby dwarf galaxies, spanning a stellar mass range of $\sim$ 10$^{7}$ to 10$^{9.5}$ $M_{\odot}$, based on MUSE wide-field mode spectroscopic observations. These galaxies are representative of ordinary star-forming dwarf galaxies in the local universe, in the sense that they follow the general stellar mass$-$metallicity relation and the star-forming main sequence relation. The metallicity gradients are compared with various galaxy properties, including stellar mass, gaseous velocity field regularity, effective yield of metals, baryonic mass, star formation rate surface density/specific star formation rate gradient, stellar gravitational potential and global/local environment, in order to probe the primary physical drivers of the metallicity distribution of dwarf galaxies. Our main results are summarized as follows:
\begin{enumerate}

    \item We find a significantly negative galaxy stellar mass$-$gaseous metallicity gradient relation (MZGR), with a best-fit slope of $-0.091\pm0.017$ dex $R_{\rm e}^{-1}$, and a Spearman's rank correlation coefficient of $-0.56\pm0.081$. The mass-dependent metallicity gradient variation is primarily driven by a higher degree of metallicity depression in the central regions of lower mass galaxies. The MZGR found here is in remarkable contrast with a lack of \citep[e.g.,][]{Lian2018} or mildly positive galaxy mass dependence of gas-phase metallicity gradient found for high mass galaxies  \citep[$>$ 10$^{9}$ $M_{\odot}$; e.g.,][]{belfiore2017}. This mass-dependent MZGR reflects the interplay of physical processes related to metal production (in-situ star formation), metal redistribution (driven by feedback, outflow or turbulence) and metallicity dilution (metal-poor gas inflow), as suggested by recent models \citep[e.g.,][]{Sharda2024}.
    
    \item Except for the primary dependence on  stellar mass, and the secondary relevant properties of gaseous velocity field regularity and effective yield (see below), all the other properties explored here have no residual correlation with metallicity gradients after controlling for stellar mass. The lack of correlation with star formation indicates that metal distribution produced by in-situ star formation is subject to substantial modulation by redistribution processes. 

    \item Galaxies with irregular gaseous velocity field are characterized by significantly disturbed ISM by stellar feedback or inflow, and are more likely to have positive metallicity gradient than those with regular velocity field, even after controlling for galaxy stellar mass. Since a negative metallicity gradient is a natural outcome from an inside-out galaxy formation process, the tendency indicates that kinematic disturbance in the ISM is accompanied by significant metal redistribution. 

    \item Effective yield of metals $\rm y_{eff}$ shows a significant, albeit moderate, negative correlation with metallicity gradient, even after controlling for galaxy stellar mass, with a partial correlation coefficient $\sim$ 0.4. $\rm y_{eff}$ accounts for 22\% of the scatter of MZGR. Moreover, a linear combination of logarithmic stellar mass and effective yield significantly enhances the correlation with metallicity gradients compared to stellar mass alone, with a correlation coefficient $r$ $=$ $-0.68$. This suggests that stellar feedback-driven outflow (as a favored explanation of the low effective yield in dwarf galaxies) plays an important role in shaping the metallicity distribution within dwarf galaxies. The above-mentioned lack of correlation with baryonic mass and stellar gravitational well indicates that feedback-driven outflow, which presumably has more profound effect on lower mass galaxies, is not the only mechanism that directly re-shape the metallicity distribution.
\end{enumerate}

Positive (i.e., inverted) metallicity gradients have been also found recently for a couple of low-mass galaxies at redshift $\gtrsim$ 2, using slitless spectroscopy \citep[e.g.,][]{Wang2019, wang2020, Wang2022b}. These high-redshift dwarf galaxies have stellar mass that falls in the high-mass end of our sample, where most nearby galaxies have negative gradients, and they are different from the local ones in the sense that they have $\sim$ $1-2$ orders of magnitude higher star formation rate (density) and are presumably experiencing much more significant gas accretion from the cosmic web or local environment. In high-redshift dwarf galaxies, the much stronger star formation results in more efficient stellar feedback-driven metallicity redistribution. However, in the local dwarf galaxies at lower masses, the generally inefficient and sporadic nature of star formation activities means that, besides stellar feedback-driven direct metal transport, other longer-term metal mixing process, such as advection and diffusion, also play important roles in building up the metallicity distribution.

\begin{acknowledgements} We acknowledge support from the National Key Research and Development Program of China (grant No. 2023YFA1608100), and from the NSFC grant (Nos. 12122303, 11973039, 11421303, 11973038, 12233008), FONDECYT Iniciaci\'on en investigaci\'on 2020 Project 11200263 and the ANID BASAL project FB210003. This work is also supported by the China Manned Space Project (Nos.CMS-CSST-2021-B02, CMS-CSST-2021-A07). We acknowledge support from the CAS Pioneer Hundred Talents Program, the Strategic Priority Research Program of Chinese Academy of Sciences (Grant No. XDB 41000000) and the Cyrus Chun Ying Tang Foundations.
\end{acknowledgements}

\bibliographystyle{bibtex/aa}
\bibliography{bibtex/reference}

\begin{appendix}
\onecolumn
\begin{landscape}
\section{Main properties of the galaxy sample}
In this section we present a table with general information and derived properties for all the galaxies in the sample. The meaning of each column is indicated in the notes below.

{\small

\setlength{\tabcolsep}{4pt}{
\renewcommand{\arraystretch}{1.2}
\begin{longtable}{llllllllllllllr}
\caption{Main properties of the galaxy sample.}\\
\hline
\label{tab:ana_pro}

Galaxy & R.A.(J2000) & Dec.(J2000) & z & D & $R_e$ &ellipticity & PA & i & log$(M_{\star})$ & log$(M_{\text{H I}})$  &$\rm y_{eff}$ & SFR  & 12 + log(O/H) & $\nabla_{r}$ [O/H] \\
- &  degrees & degrees & - & Mpc & kpc & - & degrees & degrees &log$(M_{\odot})$ & log$(M_{\odot})$ & - & log$(M_{\odot}yr^{-1})$ & - & dex/$R_{\rm e}$ \\
(1) & (2) & (3) & (4) & (5) & (6) & (7) & (8) & (9) & (10) & (11) & (12) & (13) & (14) & (15)\\
\hline
\endfirsthead
\caption{continued.}\\
\hline
Galaxy & R.A.(J2000) & Dec.(J2000) & z & D & $R_e$ & ellipticity & PA & i & log$(M_{\star})$ & log$(M_{\text{H I}})$  &$\rm y_{eff}$ & SFR & 12 + log(O/H) & $\nabla_{r}$ [O/H] \\
- &  degrees & degrees & - & Mpc & kpc & - & degrees & degrees &log$(M_{\odot})$ & log$(M_{\odot})$ & - & log$(M_{\odot}yr^{-1})$ & - & dex/$R_{\rm e}$ \\
(1) & (2) & (3) & (4) & (5) & (6) & (7) & (8) & (9) & (10) & (11) & (12) & (13) & (14) & (15)\\
\hline
\endhead
\hline
\endfoot

AGC191702      &137.15226   &5.29078       &0.001994    &12.21 &0.78  &0     &0     &0     &7.12    &7.74    &$-$2.29   &$-$2.15   &7.81    &0.115$\pm0.085$   \\
AGC193816      &140.36375   &7.36639       &0.004633    &21.47 &1.18  &0     &0     &0     &8.55    &8.33    &$-$2.43   &$-$1.45   &8.36    &$-$0.057$\pm0.036$  \\
CGCG007-025    &146.0078    &$-$0.64227    &0.004810    &23.05 &1.05  &0.501 &159.7 &62    &8.1     &-       &-         &$-$0.74   &7.68    &0.019$\pm0.01$    \\
CGCG035-007    &143.68627   &6.42563       &0.0018      &4.92  &0.41  &0.24  &63    &41    &7.6     &7.69    &$-$2.62   &$-$2.85   &7.96    &0.061$\pm0.022$   \\
ESO115-021     &39.41943    &$-$61.3517    &0.0017      &4.99  &1.06  &0.72  &41.0  &74    &8.2     &8.66    &$-$2.46   &$-$2.48   &7.82    &$-$0.12$\pm0.024$   \\
ESO119-016     &72.87165    &$-$61.65094   &0.0032      &10.08 &2.62  &0.48  &26    &58    &8.4     &7.9     &$-$2.83   &$-$2.14   &8.05    &0.088$\pm0.06$    \\
ESO158-003     &71.57227    &$-$57.34378   &0.0040      &10.20 &1.48  &0.14  &0     &30    &8.76    &6.47    &$-$2.99   &$-$1.22   &8.24    &$-$0.025$\pm0.005$  \\
ESO184-82      &293.76842   &$-$52.84389   &0.008685    &28.576&1.79  &0.305 &137.6 &47    &8.95    &8.9     &$-$2.42   &$-$0.71   &8.23    &$-$0.102$\pm0.002$  \\
ESO483-013     &63.17129    &$-$23.15887   &0.0027      &10.68 &1.36  &0.28  &125.0 &46    &8.81    &7.62    &$-$3.06   &$-$1.1    &8.11    &$-$0.058$\pm0.005$  \\
ESO486-021     &75.83203    &$-$25.42293   &0.0029      &9.11  &0.64  &0.14  &90    &30    &8.26    &8.38    &$-$2.67   &$-$1.69   &7.88    &$-$0.072$\pm0.003$  \\
FCC090         &52.78442    &$-$36.29014   &0.006048    &19.23 &1.1   &0.256 &154.5 &43    &8.95    &7.77    &$-$2.78   &$-$1.13   &8.43    &$-$0.094$\pm0.007$  \\
FCC113         &53.27854    &$-$34.80811   &0.004631    &16.14 &1.6   &0.213 &164.9 &39    &8.16    &8.16    &$-$2.52   &$-$2.13   &8.13    &0.063$\pm0.007$   \\
FCC119         &53.39101    &$-$33.57332   &0.004583    &20.14 &1.45  &0.15  &47.7  &32    &8.8     &-       &-         &$-$2.61   &8.55    &0.038$\pm0.077$   \\
FCC263         &55.38583    &$-$34.88833   &0.005751    &12.59 &1.06  &0.54  &3.6   &65    &8.71    &8.22    &$-$2.65   &$-$1.24   &8.29    &$-$0.035$\pm0.002$  \\
FCC285         &55.75914    &$-$36.27337   &0.002955    &9.16  &1.67  &0.39  &105   &54    &8.11    &8.15    &$-$2.59   &$-$1.85   &8.02    &$-$0.151$\pm0.005$  \\
FCC306         &56.43916    &$-$36.34653   &0.002955    &9.74  &0.39  &0.287 &41.1  &46    &7.61    &8.03    &$-$2.33   &$-$2.42   &7.97    &$-$0.011$\pm0.007$  \\
FCC308         &56.47854    &$-$36.35697   &0.004995    &8.68  &1.47  &0.643 &6.4   &72    &8.34    &8.38    &$-$2.19   &$-$2.07   &8.43    &$-$0.026$\pm0.003$  \\
IC1959         &53.30246    &$-$50.41425   &0.002131    &6.85  &1.27  &0.732 &149.6 &80    &8.16    &8.37    &$-$2.54   &$-$1.69   &7.93    &$-$0.054$\pm0.001$  \\
IC2828         &171.79559   &8.73105       &0.003451    &14.20 &0.7   &0.539 &60.5  &65    &8.37    &7.83    &$-$2.87   &$-$1.24   &8.03    &$-$0.011$\pm0.001$  \\
IC3476         &188.17452   &14.05044      &$-$0.00053  &13.80 &1.97  &0.312 &27.6  &48    &8.74    &8.42    &$-$2.35   &$-$0.74   &8.42    &$-$0.085$\pm0.001$  \\
IC4247         &201.68539   &$-$30.3628    &0.0014      &4.97  &0.47  &0.40  &153.0 &53    &7.74    &7.3     &$-$3.04   &$-$2.75   &7.91    &0.029$\pm0.043$   \\
IC4870         &294.40667   &$-$65.81183   &0.00292     &8.51  &1.04  &0.452 &131.4 &59    &8.49    &8.76    &$-$2.56   &$-$1.07   &7.84    &0.066$\pm0.003$   \\
MCG-03-34-002  &196.98617   &$-$16.6892    &0.0031      &7.90  &0.69  &0.38  &140.0 &52    &8.41    &7.32    &$-$3.25   &$-$2.11   &8.0     &$-$0.074$\pm0.016$  \\
NGC0059        &3.85488     &$-$21.4445    &0.0013      &5.30  &0.74  &0.30  &122.0 &47    &8.52    &6.95    &$-$3.33   &$-$1.76   &8.0     &0.037$\pm0.018$   \\
NGC853         &32.92161    &$-$9.30599    &0.005014    &21.00 &2.24  &0.450 &70.9  &59    &9.25    &8.79    &$-$2.57   &$-$0.51   &8.33    &$-$0.103$\pm0.002$  \\
NGC1311        &50.029      &$-$52.18553   &0.0019      &5.45  &0.87  &0.53  &36    &62    &8.22    &7.96    &$-$2.87   &$-$2.0    &7.96    &0.076$\pm0.011$   \\
NGC1522        &61.533      &$-$52.66842   &0.0030      &9.54  &0.84  &0.34  &37    &49    &8.53    &-       &-         &$-$1.47   &8.05    &$-$0.148$\pm0.015$  \\
NGC1705        &73.55625    &$-$53.36106   &0.00211     &5.10  &0.43  &0.286 &48.6  &46    &8.24    &8.02    &$-$2.88   &$-$1.52   &7.89    &$-$0.007$\pm0.001$  \\
NGC1796        &75.67729    &$-$61.14006   &0.003381    &10.60 &1.31  &0.526 &101.1 &64    &9.14    &8.33    &$-$2.35   &$-$0.99   &8.76    &$-$0.09$\pm0.001$   \\
NGC1800        &76.60717    &$-$31.95422   &0.00272     &8.01  &0.83  &0.524 &112.2 &64    &8.67    &8.27    &$-$2.77   &$-$1.3    &8.13    &$-$0.032$\pm0.002$  \\
NGC2915        &141.54804   &$-$76.62633   &0.00156     &4.29  &0.32  &0.422 &125   &56    &8.2     &8.28    &$-$2.6    &$-$1.77   &7.98    &$-$0.039$\pm0.001$  \\
NGC3125        &151.63905   &$-$29.93486   &0.003712    &15.00 &0.82  &0.385 &111.8 &54    &9.17    &8.62    &$-$2.72   &$-$0.18   &8.11    &$-$0.009$\pm0.001$  \\
NGC3593        &168.65417   &12.81767      &0.002075    &8.95  &2.02  &0.591 &87.5  &69    &9.64    &8.33    &$-$2.33   &$-$0.15   &8.82    &$-$0.145$\pm0.002$  \\
NGC4383        &186.35635   &16.47013      &0.005704    &17.94 &1.24  &0.427 &26.4  &57    &9.38    &9.44    &$-$2.18   &$-$0.32   &8.4     &$-$0.112$\pm0.001$  \\
NGC4592        &189.82807   &$-$0.53201    &0.003566    &11.64 &3.04  &0.410 &88.2  &55    &8.88    &9.74    &$-$1.75   &$-$0.88   &8.14    &$-$0.214$\pm0.002$  \\
NGC4809A       &193.71276   &2.65409       &0.002943    &21.90 &2.98  &0.677 &66.1  &75    &8.96    &-       &-        &$-$1.05   &7.97    &$-$0.021$\pm0.003$  \\
NGC4809B       &193.71276   &2.65409       &0.002943    &21.90 &2.56  &0.577 &160.1 &68    &9.01    &-       &-         &$-$1.08   &7.9     &$-$0.005$\pm0.002$  \\
NGC5253        &204.98318   &$-$31.64011   &0.001358    &3.55  &0.7   &0.525 &26.9  &64    &8.65    &7.97    &$-$2.86   &$-$0.81   &8.05    &$-$0.08$\pm0.001$   \\
PGC132213      &330.78995   &$-$12.37180   &0.002750    &9.31  &0.32  &0.296 &20.3  &46    &7.12    &-       &-         &$-$2.42   &7.74    &0.052$\pm0.025$   \\
UGC685         &16.8435     &16.68457      &0.0005      &4.70  &0.83  &0.18  &122   &35    &7.95    &7.84    &$-$3.03   &$-$2.18   &7.7     &0.095$\pm0.008$   \\
UGC695         &16.9435     &1.06367       &0.0022      &10.68 &1.5   &0.16  &0     &32    &8.04    &7.9     &$-$2.99   &$-$1.87   &7.75    &0.097$\pm0.029$   \\
UGC891         &20.32915    &12.41251      &0.0021      &11.10 &1.38  &0.39  &42    &52    &8.39    &8.51    &$-$2.32   &$-$2.64   &8.24    &$-$0.079$\pm0.127$  \\
UGC1056        &22.19739    &16.6884       &0.0020      &10.57 &1.06  &0.06  &0     &20    &8.32    &7.77    &$-$3.13   &$-$1.72   &7.84    &$-$0.018$\pm0.005$  \\
UGC3755        &108.465     &10.52194      &0.00105     &6.67  &1.21  &0.500 &160.3 &62    &8.05    &7.94    &$-$2.89   &$-$2.1    &7.84    &0.075$\pm0.006$   \\
UGC5288        &147.82105   &7.82783       &0.0019      &11.40 &1.48  &0.14  &151.0 &30    &8.14    &8.28    &$-$2.53   &$-$1.36   &7.97    &$-$0.093$\pm0.008$  \\
UGC5889        &161.84292   &14.06944      &0.001912    &6.89  &1.14  &0.093 &57.9  &26    &8.27    &8.14    &$-$2.52   &$-$2.65   &8.24    &$-$0.145$\pm0.014$  \\
UGC5923        &162.28139   &6.91742       &0.0024      &7.33  &0.31  &0.38  &173.0 &52    &8.24    &7.82    &$-$2.66   &$-$1.62   &8.23    &0.035$\pm0.026$   \\
UGC8041        &193.80273   &0.11665       &0.004416    &14.52 &3.43  &0.223 &173.7 &40    &9.01    &9.03    &$-$2.09   &$-$1.22   &8.55    &$-$0.203$\pm0.005$  \\
UGCA116        &88.9275     &3.39222       &0.002682    &14.43 &1.06  &0.38  &131.6 &53    &9.23    &8.92    &$-$2.64   &0.11      &8.02    &0.008$\pm0.002$   \\
UGCA193        &150.65047   &$-$6.01371    &0.0022      &9.70  &1.28  &0.73  &14    &77    &8.28    &8.19    &$-$2.57   &$-$1.97   &8.15    &$-$0.01$\pm0.005$   \\
UGCA442        &355.94034   &$-$31.9567    &0.0009      &4.27  &1.14  &0.52  &43    &62    &7.86    &8.22    &$-$2.44   &$-$2.41   &7.92    &0.041$\pm0.018$   \\
UM461          &177.88896   &$-$2.37276    &0.003465    &19.54 &0.63  &0.32  &90    &48    &7.71    &8.54    &$-$2.3    &$-$1.17   &7.58    &0.078$\pm0.046$   \\
UM462          &178.15497   &$-$2.46942    &0.003469    &19.18 &1.07  &0.102 &60.1  &27    &8.68    &8.59    &$-$2.55   &$-$0.63   &8.06    &0.036$\pm0.008$   \\
VCC0415        &185.10538   &6.90836       &0.008539    &24.20 &2.08  &0.240 &63.8  &42    &8.76    &8.37    &$-$2.61   &$-$1.19   &8.27    &$-$0.075$\pm0.004$  \\
VCC2037        &191.56375   &10.20556      &0.003809    &9.63  &1.49  &0.563 &9.6   &67    &7.56    &7.18    &$-$2.98   &$-$2.84   &7.93    &0.194$\pm0.031$   \\

\end{longtable}}
\tablefoot{Col(1):Galaxy name. (2)-(4): Celestial coordinates in J2000 and redshift of galaxy from NED Database. (5) galaxy distance from several sources: HyperLeda Database, \protect\cite{Marasco2023}, \protect\cite{Kourkchi2017} or, when not available, calculated from redshift. (6)-(9): the morphological properties as determined in this work. The zero value of ellipticity and PA means that the galaxies are too irregular or faint to be measured correctly in our program. (10) Stellar mass. (11) HI mass. Six galaxies in our sample do not have HI observations or detections are left with a single bar. (12) Effective Yield of Oxygen. (13) Star formation rate. (14) Metallicity derived from integrated MUSE spectra, using N2S2H$\alpha$ method (\protect\cite{dopita2016}). (15): Radial gradient of gas-phase metallicities. Note: NGC4809A and NGC4809B are two galaxies in the early stage of interaction in a single MUSE exposure, and we divided them into two single galaxies to measure their properties respectively.}
}
\end{landscape}
\twocolumn

\end{appendix}

\end{document}